\documentclass[acmsmall]{acmart}

\AtBeginDocument{%
  }

\setcopyright{acmlicensed}
\copyrightyear{2018}
\acmYear{2018}
\acmDOI{XXXXXXX.XXXXXXX}

\acmJournal{JACM}
\acmVolume{37}
\acmNumber{4}
\acmArticle{111}
\acmMonth{8}

\usepackage{
  color,
  float,
  epsfig,
  wrapfig,
  graphicx,
  subcaption
}
\usepackage[nolist]{acronym}
\usepackage{algorithm2e}
\usepackage{algpseudocode}
\usepackage{graphics}
\usepackage{xspace}
\usepackage{multirow, multicol}
\usepackage{colortbl}
\usepackage{hhline}
\usepackage{todonotes}
\usepackage{tabularx}
\usepackage{booktabs}
\usepackage{subcaption}
\usepackage{graphicx} 

\usepackage{
  tikz,
  pgfplots,
  pgfplotstable
}
\usepackage{hyperref}

\newcommand{\mypar}[1]{\smallskip\noindent\textbf{#1.}\xspace}

\usetikzlibrary{
  shapes.geometric,
  arrows,
  external,
  pgfplots.groupplots,
  matrix
}

\usepackage{mathtools,}

\DeclareMathAlphabet{\mathcal}{OMS}{cmsy}{m}{n}

\DeclareGraphicsExtensions{%
    .png,.PNG,%
    .pdf,.PDF,%
    .jpg,.mps,.jpeg,.jbig2,.jb2,.JPG,.JPEG,.JBIG2,.JB2}

\begin{acronym}
    \acro{ANN}{artificial neural network}
    \acro{AP}{aggregate perturbation}
    \acro{AUC}{area under the curve}
    \acro{BIM}{basic iterative method}
    \acro{CAN}{controller area network}
    \acro{CH}{car hacking}
    \acro{CNN}{convolutional neural network}
    \acro{CRC}{cyclic redundancy check}
    \acro{DL}{deep learning}
    \acro{DLC}{data length code}
    \acro{DoS}{denial of service}
    \acro{ECU}{electronic control unit}
    \acro{FFNN}{feed-forward neural network}
    \acro{FGM}{fast gradient method}
    \acro{FGSM}{fast gradient sign method}
    \acro{FN}{false negative}
    \acro{FP}{false positive}
    \acro{FPR}{false positive rate}
    \acro{GAN}{generative adversarial network}
    \acro{GPU}{graphic processing unit}
    \acro{GRU}{gated recurrent unit}
    \acro{HCRL}{hacking and countermeasures research lab}
    \acro{IDS}{intrusion detection system}
    \acro{IoT}{internet-of-things}
    \acro{IPS}{intrusion prevention system}
    \acro{LSTM}{long short-term memory}
    \acro{MAE}{mean absolute error}
    \acro{ML}{machine learning}
    \acro{MSE}{mean squared error}
    \acro{OBD}{on-board diagnostics}
    \acro{PGD}{projected gradient descent}
    \acro{READ}{reverse engineering of automotive data frames}
    \acro{RNN}{recurrent neural network}
    \acro{ROC}{receiver operator characteristic}
    \acro{RPM}{revolutions per minute}
    \acro{SVM}{support vector machine}
    \acro{TN}{true negative}
    \acro{TP}{true positive}
    \acro{TPR}{true positive rate}
    \acro{ZOO}{zeroth order optimization}
    \acro{SOF}{start of frame}
	\acro{RTR}{Remote Transmission Request}
	\acro{DLC}{Data Length Code}
	\acro{ACK}{Acknowledgement}
	\acro{EOF}{End of Frame}
	\acro{IFS}{Inteframe Space}
	\acro{IDE}{Identifier Extension Bit}
    \acro{DBC}{Database CAN}

\end{acronym}

\begin{document}


\title{Assessing the Resilience of Automotive Intrusion Detection Systems to Adversarial Manipulation}
\author{Stefano Longari}
\email{stefano.longari@polimi.it}
\orcid{0000-0002-7533-4510}
\author{Paolo Cerracchio}
\email{paolo.cerracchio@mail.polimi.it}
\author{Michele Carminati}
\email{michele.carminati@polimi.it}
\author{Stefano Zanero}
\email{stefano.zanero@polimi.it}

\affiliation{%
  \institution{Dipartimento di Elettronica, Informazione e Bioingegneria, Politecnico di Milano}
  \city{Milan}
  \state{}
  \country{Italy}
}

\renewcommand{\shortauthors}{Longari et al.}

\begin{abstract}
The security of modern vehicles has become increasingly important, with the \ac{CAN} bus serving as a critical communication backbone for various Electronic Control Units (ECUs). The absence of robust security measures in CAN, coupled with the increasing connectivity of vehicles, makes them susceptible to cyberattacks. While \acp{IDS} have been developed to counter such threats, they are not foolproof. Adversarial attacks, particularly evasion attacks, can manipulate inputs to bypass detection by \acp{IDS}.
This paper extends our previous work by investigating the feasibility and impact of gradient-based adversarial attacks performed with different degrees of knowledge against automotive \acp{IDS}. We consider three scenarios: white-box (attacker with full system knowledge), grey-box (partial system knowledge), and -- the more realistic -- black-box (no knowledge of the IDS' internal workings or data). We evaluate the effectiveness of the proposed attacks against state-of-the-art \acp{IDS} on two publicly available datasets. Additionally, we study effect of the adversarial perturbation on the attack impact and evaluate real-time feasibility by precomputing evasive payloads for timed injection based on bus traffic. Our results demonstrate that, besides attacks being challenging due to the automotive domain constraints, their effectiveness is strongly dependent on the dataset quality, the target \ac{IDS}, and the attacker's degree of knowledge.
\end{abstract}



\received{15 September 2024}
\received[revised]{1 February 2025}
\received[accepted]{18 May 2025}

\maketitle

\section{Introduction}
\label{sec:intro}

The security of modern vehicles has become an increasingly critical concern as the integration of connected technologies, such as Bluetooth and 5G, exposes vehicles to potential cyber threats. The \ac{CAN} bus serves as the communication backbone for various \acp{ECU}, managing essential vehicle functions ranging from engine control to infotainment. However, the CAN protocol lacks fundamental security features such as encryption and authentication, leaving it vulnerable to cyberattacks. This issue was prominently highlighted in 2015 by Miller and Valasek, who remotely attacked a Jeep Cherokee~\cite{miller2015remote,Miller2016}.
\acp{IDS} have been widely adopted as a key defense mechanism to monitor network traffic and identify malicious activities, mainly through frequency-based detectors, analyzing patterns like packet timing or sequence, and payload-based detectors, examining packet content for irregularities. Traditionally, \acp{IDS} rely on machine learning models to detect anomalies~\cite{rajapaksha2023ai,longari2020cannolo}, but these models are susceptible to adversarial attacks, wherein carefully crafted inputs evade detection~\cite{Goodfellow2013}. Current research on adversarial threats, however, mainly focuses on computer vision, with relatively few specialized studies in the automotive sector~\cite{cerracchio2024investigating, longari2023evaluating}. 

In our previous work~\cite{cerracchio2024investigating}, we explored the resilience of automotive \acp{IDS} against white-box and grey-box adversarial attacks, where the attacker has either full or partial knowledge of the target system and data. The evasion attacks leverage gradient-based techniques, algorithms that were inspired from the computer vision domain but adapted to the automotive one, to craft adversarial examples that bypass \ac{IDS} detection. 
However, in a real-world setting, attackers often lack direct access to the internal workings of the \ac{IDS}, making black-box attacks a more realistic threat. Black-box attacks occur when the adversary has no knowledge of the IDS's architecture, parameters, or training data, and must rely either on probing the original system or on a surrogate model to infer attack strategies. Thanks to the transferability property of gradient-based Techniques~\cite{szegedy2013intriguing}, we focus on the ability of adversarial examples generated by surrogate models to transfer to grey- and black-box target IDSs.
Given this premise, this paper extends our prior work by investigating the feasibility and impact of white-, grey-, and black-box adversarial attacks on automotive \acp{IDS}.

Our experimental evaluation on the ReCAN~\cite{zago2020recan} and CARHacking~\cite{seo2018gids} datasets provides a comprehensive evaluation of adversarial robustness of automotive \acp{IDS}.
Our objective is to identify the effectiveness, impact, and feasibility of adversarial attacks in the automotive domain. Through our experiments, we first explore the vulnerability of \ac{IDS} models in white- grey- and black-box scenarios, identifying the capability of adversarial algorithms to evade them. Then, we study the impact of adversarial perturbations on the network signals, to evaluate whether the meaning of the attack is maintained after the perturbation. Finally, we evaluate the feasibility of executing such attacks in real-time constrained systems by precomputing evasive attack sequences and identifying injection points in the CAN data stream.  
Our results demonstrate that the attack's effectiveness is strongly dependent on the dataset quality, the target \ac{IDS}, and the attacker's degree of knowledge. 
As expected, white-box attacks proved overall most effective, degrading \ac{IDS} performance by up to 60\% on the most challenging dataset. Nonetheless, the impact of adversarial samples in grey- and black-box scenarios achieve lower but impactful results of up to 39\% and 43\% respectively. 
Interestingly, in the grey- and black-box scenarios, perturbation sometimes led to an increased detection rate, suggesting that in such constrained knowledge scenarios, the attempted perturbations may make the original attack more conspicuous to the target \ac{IDS}. 
These aspects suggests that knowledge of the target \ac{IDS} and the quality of data are crucial. The various models perform very differently, with autoencoder-based IDSs being more resistant to adversarial attacks, but predictor-based oracles being the more effective at generating adversarial samples. We conclude by discussing mitigations and the challenges to face in order to apply them.
In summary, our contributions are the following:

\begin{itemize}

\item We investigate the feasibility and impact of adversarial attacks on payload-based automotive \acp{IDS} and different datasets, considering an attacker with different degrees of knowledge. 

\item We explore the transferability of adversarial examples generated by substitute models to grey- and black-box target IDS, addressing a more realistic and challenging scenario.

\item We evaluate the impact of adversarial perturbations on the actual vehicle signals, providing a qualitative assessment of the attack's effectiveness in the different scenarios. 

\end{itemize}

\section{Background on CAN Security}
\label{sec:background}

\mypar{CAN Primer} 
The \acf{CAN} is a multi-master, message-broadcast protocol originally developed by Robert Bosch in 1986, and later updated and standardized in 2012 (ISO 11898-1). It is one of the most widely adopted network protocols and has become the de-facto standard for vehicle onboard networks, primarily used in the automotive industry to connect various \acp{ECU} within a vehicle. The CAN bus~\cite{can-cia} operates as a broadcast medium, offering multi-master capabilities. This allows any node connected to the bus to read any packet transmitted across the network. Additionally, when the bus is idle, any node can transmit a message. If multiple nodes attempt to communicate simultaneously, message arbitration is resolved by prioritizing the message with the lowest ID. 
The \ac{CAN} protocol incorporates various error checking mechanisms. If a message incurrs in an error it generates an error frame and invalidates the packet. Should errors persist, the malfunctioning node eventually removes itself from the network~\cite{can-texasinstruments}.

Four types of messages can be transmitted on a \ac{CAN} network: \textit{Data}, \textit{Remote}, \textit{Error}, and \textit{Overload} frames. The \textit{Data} frame is the most prevalent and, as its name implies, is used to transmit payload data. \textit{Remote} and \textit{Overload} frames are less frequently encountered in modern \ac{CAN} systems. 

\textit{Error} frames are sent by either the transmitter or the receiver when an error is detected in the packet being transmitted on the bus, signaling that the current packet is invalid, and are not usually notified by the CAN controller to higher layer computation units. 

\textit{Data} frames payloads embed multiple values each, each with a distinct meaning. In this paper, we refer to these values as signals. Signals typically convey information from sensors or commands for actuators, but they can also contain noisy bit sequences or \acp{CRC}, complicating their interpretation for security researchers. The mapping of these signals to specific \ac{CAN} IDs is often proprietary information of the manufacturers, which do not disclose them. This reliance on security through obscurity presents significant challenges in the design of \acp{IDS} for \ac{CAN}.

\mypar{CAN Security Issues} \label{sec} \ac{CAN}, originally designed with network isolation in mind, lacks any inherent security mechanisms and is thus vulnerable to various security threats~\cite{cansecurityandvulnerabilities}.
Lacking \textit{authentication} and \textit{encryption}, and being a broadcast network, any node can intercept messages and perform actions aimed at compromising the authenticity and integrity of the transmitted data. Such as transmit spoofed messages with manipulated \ac{CAN} IDs. Moreover, due to \ac{CAN}’s arbitration system, which prioritizes messages based on \ac{CAN} IDs, an attacker can \textit{misuse the protocol} by sending messages with low \ac{CAN} IDs (higher priority), thus preventing legitimate messages from winning arbitration and causing a \ac{DoS} attack. Recent approaches to \ac{CAN} attacks focus triggering the error-handling mechanisms of CAN to silence specific nodes. This results in a targeted \ac{DoS} attack where only the victim node is excluded from the bus~\cite{tron022canflict}.

\mypar{ML-based Intrusion Detection in CAN} Intrusion detection is a widely studied field~\cite{denning1987intrusion}. For a comprehensive review of automotive IDSs, refer to~\cite{lampe2023survey}. This section provides a concise overview of key concepts. 
Most CAN intrusion detection systems focus on anomaly detection, which identifies deviations from normal behavior. This classification can effectively differentiate between legitimate and injected messages in a network. \ac{DL} techniques often use autoencoders~\cite{Chen2018} or predictive models~\cite{taylor2016anomaly}. Autoencoders learn a compressed representation of input data, then attempt to reconstruct it, while predictive models forecast the next sequence element based on previous samples. Both approaches operate in an unsupervised framework, assuming deviations from the training data represent anomalies. The difference between the reconstructed or predicted sample and the actual target, known as the \emph{anomaly score}, quantifies deviation from the learned distribution. A threshold is used to classify instances as anomalous or not, typically derived from errors in a dedicated non-anomalous thresholding set.

CAN \acp{IDS} can be flow-based, payload-based, or hybrid~\cite{nichelinipozzoli}. Flow-based detection exploits deterministic ECU packet patterns. Taylor et al.\cite{Taylor2015} show that an \ac{SVM} can detect malicious outliers using metrics such as frame count, average inter-arrival time, and Hamming distance. However, frequency-based methods struggle with impersonation attacks that mimic normal ECU behavior while manipulating data fields. Time-based models also yield comparable results\cite{moore2017modeling,tomlinson2018detection}. Flow-based \acp{CNN} detectors exist but require matrix-structured inputs, increasing complexity and vulnerability to adversarial attacks~\cite{hossain2020effective,song2020vehicle,lampe2023survey}.

Payload-based \acp{IDS} address these limitations by detecting subtle impersonation attacks but are generally more complex. Early works~\cite{kang2016intrusion,chockalingam2016detecting,taylor2016anomaly} use 64-bit data fields as input for \ac{ML} models. Taylor et al.\cite{taylor2016anomaly} propose a predictive \ac{LSTM} that performs well on synthetic attacks, while Tanksale\cite{Tanksale2020} focuses on single signals in the data flow, though identifying fields like \ac{RPM} and brake position remains difficult due to proprietary obfuscation. CANet~\cite{hanselmann2020canet} employs an \ac{LSTM}-based autoencoder with a separate interface for each CAN ID, concatenating outputs for final reconstruction via an \ac{FFNN}, leading to a complex model. CANnolo and CANdito~\cite{longari2022candito,longari2020cannolo} simplify this by reconstructing payload windows for individual IDs, expanding on the \ac{READ} method~\cite{MarchettiStabili}.

 \section{Related Work}
\label{sec:relwork}
In this section, we focus on relevant works in the field of adversarial machine learning.

Adversarial machine learning is a branch of \ac{ML} that focuses on studying attacks against \ac{ML} algorithms. Goodfellow et al.\cite{Goodfellow2013} demonstrated that small perturbations applied to the input of a deep learning model can induce classification errors in many realistic settings. These specially crafted inputs are known as \emph{adversarial examples}\cite{Goodfellow2014}. Such techniques can be exploited by malicious attackers in various ways, each with different objectives~\cite{chakraborty2018adversarial}: \textbf{(a)} Exploratory attacks involve probing a model that behaves as a black box to extract knowledge based on its responses to different inputs. \textbf{(b)} Evasion attacks are the most common and were the first to be studied; here, the adversary manipulates inputs to cause misclassification. \textbf{(c)} Poisoning attacks involve contaminating the training data, which in many real-world scenarios requires evading checks by human experts or by earlier versions of the targeted system, leading to mislabeling when such data is used in training.

This research field, originally stemming from computer vision, has also been applied to security-critical applications, such as network or transaction monitoring and malware detection~\cite{Lambert2020}. Previous research has revealed interesting properties of adversarial examples: Papernot et al.\cite{papernot2017practical} demonstrated the transferability of adversarial attacks by training an \emph{oracle}—a substitute network—to carry out attacks instead of directly targeting the model. Longari et al.\cite{longari2023evaluating} designed a similar oracle-based approach within the domain of automotive \acp{IDS}, developing a greedy algorithm for black-box adversarial attacks. However, this work evaluates the distortion introduced in the perturbed packets using Hamming distance, which does not fully capture the actual semantic distance and overlooks attackers with varying levels of knowledge about the target system.

One immediate way to enhance the resilience of \ac{DL} models in supervised or semi-supervised settings is adversarial training, which involves including labeled adversarial examples in the training set~\cite{wong2020fast}. Another approach is to select more resilient input features; for example, Papernot et al.~\cite{papernot2016distillation} propose training an initial model and then approximating it with a second, more resilient model, leveraging the confidence scores and class similarity insights from the first model.

In the context of network intrusion detection, Li et al.\cite{li2021adversarial} attacked an in-vehicle Ethernet monitored by an \ac{LSTM} \ac{IDS} classifier\cite{khan2019long} using \ac{FGSM} and \ac{BIM}, achieving a recall score as low as $2\%$. The authors then retrained the \ac{LSTM} with adversarial examples, nearly restoring the baseline attack-free score ($\sim98$\%). Similarly, Sauka et al.~\cite{sauka2022adversarial} conducted multiple tests against \ac{FGSM}, \ac{PGD}, and successfully mitigated the attack using adversarial training.

\section{Threat Model}
\label{sec:threat_modeling}

As highlighted in prior research~\cite{nichelinipozzoli,lampe2023survey}, the specific objectives of attacks on the CAN bus can vary, but they generally involve injecting sequences of packets to achieve outcomes ranging from impairing vehicle functionality to compromising the safety of passengers and nearby individuals. These attacks typically work by introducing false sensor data or forged control commands.
Given the lack of inherent security mechanisms in the CAN protocol, as discussed in Section~\ref{sec:background}, we assume an attacker with control over a node in the network, which is plausible given the established methods for carrying out complex attacks on CAN systems. 
This level of control allows the attacker to inject or remove packets to obtain its goals.

\begin{table}
\centering
\caption{Overview of attacker knowledge and capabilities across threat scenarios}
\label{tab:attacker_scenarios}
\resizebox{.8\textwidth}{!}{
\begin{tabular}{|l|c|c|c|}
\hline
 & \textbf{White-box} & \textbf{Grey-box} & \textbf{Black-box} \\
\hline
\textbf{Access to IDS model architecture} & Full & None & None \\
\hline
\textbf{Access to IDS training data} & Full & Full & None \\
\hline
\textbf{Oracle} & Not required & Required & Required \\
\hline
\textbf{Training dataset for oracle} & - & Same as IDS & Similar but independent \\
\hline
\end{tabular}
}
\vspace{-10px}
\end{table}

We define an \textbf{evasive attack} as one in which the injected packets maintain their malicious intent (e.g., spoofing sensor values or control commands) while being purposefully crafted to avoid detection by the IDS. In other words, our adversarial strategy focuses on perturbing a predetermined set of malicious packet sequences with the goal of transforming them into evasive examples. The attacker then - controlling a node on the bus - injects the sequence of perturbed packets on the bus. To perturb them effectively, an attacker may have black-, grey-, or white-box knowledge of the target IDS, each with varying degrees of access and requirements. In Table~\ref{tab:attacker_scenarios} we provide a summary of the core properties of each attacker, of which a more detailed explanation follows:

\mypar{Black-box Attacker}
The attacker has no access to the internal architecture or training data of the target \ac{IDS}. To generate adversarial inputs, they rely on a surrogate (oracle) model~\cite{papernot2017practical} trained to approximate the target IDS’s behavior. This surrogate is trained on a separate but compatible dataset, which could realistically be collected from a similar vehicle—e.g., one of the same make and model. The generated adversarial examples are then transferred with the goal of evading detection by the target IDS. In our experiments, this scenario is simulated by splitting the dataset into two disjoint subsets: one used to train the surrogate model and the other for the target IDS.

\mypar{Grey-box Attacker}
In the grey-box scenario, the attacker has access to the training dataset used by the target \ac{IDS}, but no knowledge of its internal architecture or parameters. This access could be obtained through a data leak or breach, though the specific method is beyond the scope of this work. The attacker uses this dataset to train a surrogate model that approximates the target’s decision boundary, and generates adversarial samples through the surrogate model.

\mypar{White-box Attacker}
The attacker has full knowledge of the target \ac{IDS}, including its model architecture, parameters, training data, and expected behavior of the monitored node. This privileged access allows them to accurately predict traffic from the victim ECU and craft coherent, stealthy attack sequences that blend into the observed communication. Such a scenario typically assumes insider access or deep system compromise.

\section{Motivation}
\label{sec:motivation}

In the automotive domain, machine learning models are already being used and have been proposed for tasks that directly affect vehicle safety, efficiency, and security. Ensuring the security of these models is therefore critically important. Within this context, Adversarial Machine Learning (AML) attacks pose a unique threat due to their ability to algorithmically generate evasive examples designed specifically to bypass detection mechanisms.

To provide a complete assessment of evasion feasibility in automotive domains, unlike previous studies where adversarial samples are evaluated in ideal scenarios, this work emphasizes creation of samples coherent with the concurrent non-anomalous network traffic. Additionally, we aim to preserve characteristics typical of automotive attacks, which often do not involve a single packet injection but rather sequences of harmful packets, while attempting to minimize deviation from the original attack sequence.
Specifically, we evaluate the effectiveness, feasibility, and impact of these attacks in the automotive context by addressing three key aspects: (i) their ability to evade state-of-the-art intrusion detection systems, (ii) their capacity to preserve the intended malicious semantics, and (iii) the possibility of overcoming the real-time constraints of the CAN bus through the precomputation of evasive attack sequences.

\mypar{Our previous work} The foundation for this research is laid by our previous work, Cerracchio et al.~\cite{cerracchio2024investigating}, which began investigating the impact of evasion attacks on automotive \acp{IDS}. We adapted well-known gradient-based adversarial techniques from the computer vision domain to the automotive context, focusing on payload-based \acp{IDS} that leverage machine learning models. Our findings highlighted the feasibility of such attacks and emphasized the role of model complexity and attack quality in determining the success of evasion attempts. 
Building on these results, the present work extends the investigation by exploring three attacker scenarios, white-box, grey-box, and black-box, integrating the latter, which is the most realistic attack setting, where adversarial samples are crafted using a surrogate detection model trained on data from a similar vehicle, and evaluated on the target IDS.
To simulate varying degrees of attacker knowledge, we re-executed all experiments using a reduced dataset\footnote{The original datasets were split into two sets each, simulating scenarios with different levels of attacker knowledge.}, to provide consistent results across the various attacker models. Additionally, we replicated the experiments on a second dataset to assess how dataset variability influences adversarial behavior. The combination of an additional dataset and a systematic exploration of all three attacker scenarios allows us to draw clearer conclusions about the feasibility of adversarial attacks on CAN-based IDSs. Finally, we integrate the work with a discussion on countermeasures.


\section{Approach}
\label{sec:approach}
The objective of our study is to investigate how evasion attacks can be adapted to the domain of automotive intrusion detection and to assess their effectiveness. In doing so, we aim to identify attack strategies that are compatible with the unique characteristics of this domain — namely, the tabular and temporal nature of the data, the sensor-like behavior of certain signals within the payloads, and the real-time constraints imposed by the CAN bus environment.
Our approach consists of two key steps: first, we perform a domain-specific preprocessing phase that transforms raw CAN payloads into a representation suitable for machine learning-based IDSs. Second, we adapt and extend state-of-the-art gradient-based adversarial techniques—originally developed for computer vision—to align with the constraints of automotive systems. 
These adaptations include limiting perturbations to physically plausible ranges, preserving signal semantics, and lowering the average convergence iterations to increase compatibility with real-time execution. 
We opt for gradient-based approaches due to their transferability property~\cite{szegedy2013intriguing}: being it unlikely for an attacker to possess the actual models running on the target IDS, the capability of gradient-based attacks to transfer to other models comes helpful to allow for the training of the attacks against an oracle and their execution against a different target model.

\begin{figure*}[t]
    \centering
    \includegraphics[width=0.8\textwidth]{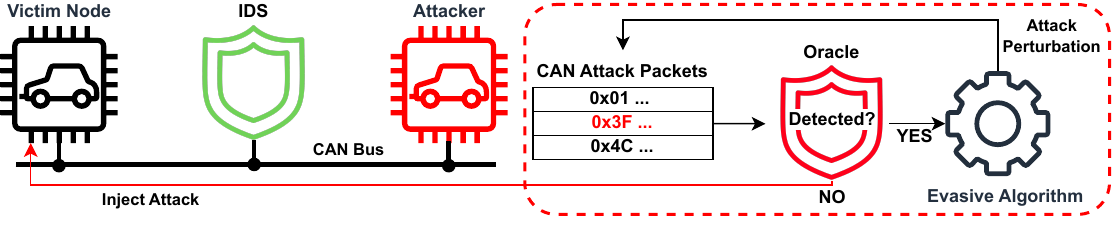}
    \caption{Scheme of the attacker's approach. The attacker controls a node with bus access. Either in real-time or precomputing it, they generate the evasive attack sequence through the oracle and inject it on the CAN bus. }\label{fig:attack}
    \vspace{-0.3cm}
\end{figure*}

\begin{table}
        \centering
        \caption{Bit ranges with semantic meaning identified}
        \label{tbl:bit_ranges}
        \resizebox{\textwidth}{!}{
        \small
        \begin{tabularx}{\columnwidth}{|p{1.6cm}|X|}
        \hline
        \textbf{Type} & \textbf{Description} \\
        \hline
        Constant & Bits that remain constant. They are excluded from the feature vector. \\
        \hline
        Physval & Sequences of bits that contain a changing value, usually corresponding to physical signals. \\
        \hline
        Binary & Ranges containing an isolated, non-constant bit, interpreted as a logical flag. \\
        \hline
        CRC & Checksums for message integrity, detected through their Gaussian random distribution. \\
        \hline
        Counter & Application-level counters, increasing by one with each subsequent frame. \\
        \hline
        \end{tabularx}
        }
\vspace{-10px}
    \end{table}

\mypar{Preprocessing} 
To ensure domain-relevant input representations, we perform a preprocessing step that extracts semantically significant features from raw payloads, addressing the lack of publicly available signal mappings and the obfuscated nature of CAN data. Following the approach proposed in prior works~\cite{zago2020recan,longari2020cannolo}, we apply a heuristic analysis to the packet payloads, incorporating slight improvements over the original \ac{READ} method~\cite{MarchettiStabili}. This analysis enables the identification of bit ranges with distinct semantic roles by examining how frequently bits change over time. For example, it distinguishes adjacent bits that function as counters from isolated binary flags that do not exhibit strong correlation with neighboring bits. Using attack-free data, we categorize the extracted fields into five types, as summarized in Table~\ref{tbl:bit_ranges}. Consistent with findings from previous work~\cite{longari2022candito}, we retain only physical and binary ranges as features, which improves both reconstruction accuracy and model efficiency compared to using the full set of non-constant bits. Finally, physical signals are parsed as integers and normalized in the $[0,1]$ interval by dividing each value by the maximum representable number for its corresponding bit length.

\mypar{Evasion Algorithms}
In our approach~\cite{cerracchio2024investigating}, as visible in Figure~\ref{fig:attack}, the attacker starts with an initial CAN log containing some unperturbed attack messages as a baseline, and attepmts to morph these sequences to be evasive by applying repeated modifications. 
We formalize an evasion attack as the optimization problem of finding the adversarial sample $\tilde{x}$, undetected under the discrimination function $F(x)$ while minimizing the perturbation $\delta (\tilde{x},x)$:
    $    \tilde{x} = argmin_{x^*}[\delta(x^*,x)]~\quad \textrm{s.t.}~ F(x^*)=0 \land x^*\in D_x$.
In most applications, the generated adversarial sample $\tilde{x}$ must also comply with domain-specific constraints, i.e., $\tilde{x} \in D_x$. For instance, in the computer vision domain, inputs are typically constrained to pixel intensity values within the integer range $[0, 255]$. 
In our case, we enforce domain validity by restricting each perturbed, normalized signal obtained during the preprocessing stage to remain within the $[0, 1]$ interval.

The algorithms we propose are derived from gradient-based techniques originally designed for computer vision; the common rationale behind these kinds of techniques, namely \ac{FGM}, \ac{BIM}~\cite{goodfellow2018BIM}, and Deepfool~\cite{moosavi2016deepfool}, is to leverage the backpropagation algorithm against the model under attack. The intuition is to push the original input towards areas in the problem space with lower confidence, approaching and crossing the decision boundary by exploiting the gradient ascent of an arbitrary loss function. 
Figure~\ref{fig:attack} illustrates a single iteration of the process. At each iteration, the chosen algorithms try to find an optimal additive term and produce $x^{t+1}$.

\subsection{\ac{BIM}-based algorithms} 

\RestyleAlgo{ruled}
\begin{algorithm}
\footnotesize
    \caption{Proposed \ac{BIM} step decay variant}\label{alg:bimdecay}
    \Begin{
            $t \gets 0$\;
            $step \gets \epsilon$\;
            \While{$t<max\_iter$}{
                $score \gets \texttt{get\_score}(ids\_model, sample)$\;
                $grad \gets \texttt{gradient}_{sample}(score)$\;
                $pert \gets step \cdot sign(grad)$ \tcp*[r]{\eqref{eq:fgsm}}
                $pert \gets -pert \cdot tamper\_mask$ \tcp*[r]{minimizing}
                $sample \gets \texttt{clip}_{min,max}(sample + pert)$\;
                $sample \gets \texttt{round}(sample)$\;
                $step \gets step \cdot decay$ \tcp*[r]{step decay}
                \If{$\texttt{get\_score}(ids\_model, sample)<threshold$}{
                    \textbf{return} $sample$ \tcp*[r]{sample is now evasive}
                }
                $t \gets t+1$\;
            }
            \Return{$None$} \tcp*[r]{Abort computation}
    }
\end{algorithm}

\begin{algorithm}
\footnotesize
    \caption{Proposed  $l2$ \ac{BIM} variant}\label{alg:biml2}
    \Begin{
            $t \gets 0$\;
            $step \gets \epsilon$\;
            \While{$t<max\_iter$}{
                $score \gets \texttt{get\_score}(ids\_model, sample)$\;
                $grad \gets \texttt{gradient}_{sample}(score)$\;
                $pert \gets step \cdot\frac{grad}{\lvert\lvert grad \rvert\rvert_2}$ \tcp*[r]{\eqref{eq:l2fgm}}
                $pert \gets -pert \cdot tamper\_mask$ \tcp*[r]{minimizing}
                $sample \gets \texttt{clip}_{min,max}(sample + pert)$\;
                $sample \gets \texttt{round}(sample)$\;
                \If{$\texttt{get\_score}(ids\_model, sample)<threshold$}{
                    \textbf{return} $sample$ \tcp*[r]{sample is now evasive}
                }
                $t \gets t+1$\;
            }
            \Return{$None$} \tcp*[r]{Abort computation}
    }
\end{algorithm}
The original \ac{BIM} iteratively applies the \ac{FGSM} perturbation described by~\eqref{eq:fgsm}. 

\begin{equation}
\footnotesize
    x^{t+1} = clip_{X}(x^t - \epsilon \frac{\nabla_x L(w,x^t)}{\lvert\lvert \nabla_x L(w,x^t) \rvert\rvert})
    \label{eq:fgsm}
\end{equation}

From a geometric point of view, this method produces a perturbation directed towards a maximum of the loss function $L$ via approximation of the gradient $\nabla_x L$ and constrained by $\lvert\lvert x^{t+1}-x^t\rvert\rvert_\infty\le\epsilon$, depending on the hyperparameter $\epsilon$.

We implement two slightly different variants of the \ac{BIM} attack, namely the \emph{step decay} \ac{BIM} and the $l2$ \ac{BIM}; both versions of the algorithm include an adjustment procedure to restrict the resulting value $x^{t+1}$ to stay in the problem space. We achieve this by clipping and rounding to the nearest integer so that the features can be effectively represented in the actual underlying bit vector. This ensures that modified payloads remain within valid signal representations, aligning with the bit-level constraints and signal encoding formats typical of in-vehicle CAN traffic. Besides this modification, the \emph{step decay} method differs from the simple \ac{BIM} in the parameter $\epsilon$: in our implementation, it has not a fixed magnitude but rather a geometrically decreasing value according to the update rule $\epsilon^{t+1}=\epsilon^{t}*\omega$, introducing the hyperparameters $\epsilon_0$ and $\omega$. The reason for the introduction of \emph{step decay} is that, similarly to what happens with the learning rate decay during the training of \ac{DL} models~\cite{you2019does}, 
the introduction of a geometrically decreasing step size helps mitigate oscillations around the decision boundary, reducing the number of iterations needed and making the attack more compatible with the strict real-time constraints of automotive systems.
\begin{equation}
\footnotesize
    x^{t+1} = clip_{X}(x^t - \epsilon \frac{\nabla_x L(w,x^t)}{\lvert\lvert \nabla_x L(w,x^t) \rvert\rvert_{2}})\label{eq:l2fgm}
\end{equation}

Conversely, the $l2$ \ac{BIM} simply uses the Euclidean norm instead of the absolute value for the \ac{FGSM} equation, resulting in \eqref{eq:l2fgm}. In this case, $\epsilon$ quantifies the module of the perturbation vector, which now completely orients itself according to the gradient and constitutes the tighter bound $\lvert\lvert x^{t+1}-x^t\rvert\rvert_2\le\epsilon$. 
We deem this approach more suitable in our case - given the higher dimensionality and feature inter-correlation in the automotive domain - but also in general when dealing with diverse features, as they apply small perturbations along all dimensions.
By relying on the Euclidean norm, the $l_2$ BIM distributes the perturbation more evenly across all features—an approach particularly well-suited for the highly interdependent and heterogeneous signals found in automotive payloads. 
In both implementations, we choose as loss function the opposite of the anomaly score.

\begin{algorithm}[t]
\footnotesize
    \caption{Pseudocode for the  DeepFool variant}\label{alg:df}
        \Begin{
        $t \gets 0$\;
        $step \gets \epsilon$\;
        \While{$t<max\_iter$}{
            $score \gets \texttt{get\_score}(ids\_model, sample) - threshold$\;
            $grad \gets \texttt{gradient}_{sample}(score)$\;
            $pert \gets \frac{score \cdot grad}{\lvert\lvert grad \rvert\rvert_2^2}$ \tcp*[r]{\eqref{eq:df}}
            $pert \gets (1+\epsilon)pert \cdot tamper\_mask$ \tcp*[r]{projection}
            $sample \gets \texttt{clip}_{min,max}(sample + pert)$\;
            $sample \gets \texttt{round}(sample)$\;
            \If{$\texttt{get\_score}(ids\_model, sample)<0$}{
                \Return{$sample$} \tcp*[r]{sample is now evasive}
            }
            $t \gets t+1$\;
        }
        \Return{$None$} \tcp*[r]{Abort computation}
\vspace{-10px}
    }
\end{algorithm}
\subsection{DeepFool-based algorithm} 
Deepfool~\cite{moosavi2016deepfool} is another iterative method that approximates the \ac{IDS} as an affine classifier $w\cdot x + b = F(w,x)$, then the perturbation $\delta$ tries to push the sample beyond the affine decision boundary, which we assume to be the 0 plane for $F$.
\begin{equation}
\footnotesize
    x^{t+1} = x^t + (1+\epsilon)\cdot(- F(w,x) \frac{\nabla_x F(w,x^t)}{\lvert\lvert \nabla_x F(w,x^t)\rvert\rvert_{2}^2}))\label{eq:df}
\end{equation}

Equation~\eqref{eq:df} illustrates the original computation, notice the overshooting factor $1+\epsilon$: since the algorithm can't converge to a point precisely on the decision boundary, we attempt to cross it by a small value $\epsilon$ instead - according to the approximation - and cause the objective misclassification. 
In our case, we do not deal with a sign-dependent decision value like $F(w,x)$, however, we can consider the reconstruction error $L(w,x)$ as a classification confidence score and apply a shift by the thresholding value $\theta$ to obtain an analogous zero-centered boundary $F(w,x) = L(w,x) - \theta$. We apply the same adjustment procedure to obtain valid samples that we use for the \ac{BIM} algorithms. This may hinder the algorithm convergence; however, we mitigate this phenomenon by testing different overshooting magnitudes. According to the evaluation of the original paper, albeit more computationally complex, it should terminate in fewer iterations than simple \ac{BIM}.
Note that in most automotive intrusion detection algorithms, the input windows contain malicious and legitimate packets. This is true for all the models in our experimental evaluation except the \ac{FFNN}. Therefore, we apply the computed perturbation only on the injected packages at each step via a simple projection of the computed perturbation matrix. Notably, in this way the algorithm naturally aligns with the temporal nature of predictive IDSs: the attacker perturbs the most recent packet to better match the model’s prediction, and upon successful evasion, the sliding input window incorporates this crafted packet, subsequently affecting future classifications.

\section{Experimental Evaluation}
\label{sec:eval}

In this section, we describe the experiments and discuss the results that are the core of our investigation, which  aim to answer the following research question: 

\textit{Are existing adversarial evasion attacks effective against payload-based automotive IDSs?}

\noindent The \textit{first experiment} evaluates the feasibility and impact of adversarial attacks against \acp{IDS} performed by an attacker with different degrees of knowledge (see Section~\ref{sec:threat_modeling}) of the target system, using the ReCAN~\cite{zago2020recan} and CarHacking~\cite{seo2018gids} datasets.

\noindent The \textit{second experiment} evaluates whether the adversarially perturbed attacks maintain the attack goal, assessing the effectiveness of the adversarially perturbed attacks on the vehicle by comparing the shape of the original and adversarial attacks signals.

\noindent Finally, the \textit{third experiment} is meant to study whether attacks can be executed even in stringent real time constraints, by restricting the assumption of the attacker's computational capabilities. To emulate the physical real-time constraints, we precompute the evasive payload sequence allowing an attacker to inject it entirely when given requirements on bus traffic are met. 

\subsection{Experimental Settings} 
\label{ssec:experimentalsettings}

\mypar{Performance Metrics} 
Since the goal of this work is to assess the impact of adversarial evasion attacks across different IDS architectures, the attacker does not alter benign traffic. As a result, the IDSs’ behavior on legitimate (non-malicious) traffic remains unchanged. Consequently, metrics such as true negatives (TN) and false positives (FP) are unaffected by the attack and are therefore not relevant to our evaluation.
We mainly report the \acf{TPR} (i.e., recall) of each IDS under attack, as it directly reflects the degradation in detection capability due to adversarial perturbations. 
Moreover, to quantify the magnitude of the adversarial perturbation applied to evade detection, we introduce the \emph{\ac{AP} metric}, which measures the \emph{mean maximum perturbation of a single field} in each packet. Formally, we define it as:
\[
\texttt{AP} = \frac{\sum_{i=1}^{N} \lvert\lvert x_{i}-\tilde{x}_{i} \rvert\rvert_{\infty}}{N}
\]
where $N$ is the number of malicious packets in the test set, $x_{i}$ is the array of features in the original $i$-th malicious packet, and $\tilde{x}_{i}$ is the corresponding adversarial packet. The infinity norm captures the largest individual feature perturbation per packet. Since all features are normalized to the $[0,1]$ range, the AP score also lies within this interval.

\subsection{Datasets Under Analysis} 
\label{sec:dataset}

For our evaluation we rely on two publicly available, real-world traffic datasets: ReCAN~\cite{zago2020recan} and the Car Hacking Dataset (CH)~\cite{seo2018gids}. Both datasets offer realism in the untampered section of the data, being extracted live from real vehicles. Where the ReCAN dataset integrates synthetic attacks to allow for complex attack events to be tested, the CH dataset contains real-world attacks, which are however executed live with safety in mind and therefore less complex and predictable.

\subsubsection{ReCAN Dataset}

The \emph{ReCAN dataset}~\cite{zago2020recan} consists of real CAN traffic collected from a Giulia Veloce vehicle using a CANtact interface~\cite{cantact}. We focus on the ``C-1'' subset, which provides the richest trace (over two hours of city and highway driving). This dataset contains no attack instances, enabling us to inject \emph{synthetic} attacks in a controlled and isolated manner, in line with prior work~\cite{longari2020cannolo,longari2022candito}.

Synthetic attacks are generated using the \emph{CANtack} tool and following the same setup presented by Nichelini et al.~\cite{nichelinipozzoli}, which supports both injection and masquerade strategies. Specifically, the following attack types are included:

\noindent\textbf{Injection Replay}: Injects previously observed packets at a lower rate (injection rate = 0.4) to avoid triggering frequency-based detection.

\noindent\textbf{Masquerade Replay}: Simulates ECU impersonation, replaces 25 valid packets content with content of a previously recorded sequence of valid packets, without affecting frequency or period of arrival. 

\noindent\textbf{Continuous Change}: Replaces 25 legitimate packets content without affecting frequency or period of arrival, but instead of replaying previous data it gradually modifies a single signal (of at least 9 bits) over 25 packets to reach a random target value inside the boundaries of the signal.
    
\noindent\textbf{Change to Minimum}: A variant of the above, targeting a final all-zero value.

\noindent\textbf{Fuzzy}: Randomly modifies physical and binary fields at each packet.

In the context of CAN security, injection attacks typically involve the insertion of new packets by the attacker, superimposed on the existing traffic. This often results in a noticeable change in the arrival frequency of the attacked packets, which could be detected. In contrast, masquerade attacks usually refer to silencing the sender ECU by various methods~\cite{tron022canflict,Miller2016}, and then sending packets on its behalf at the correct frequency to avoid detection by frequency-based IDSs. To further hide their presence, attackers may execute attacks that do not abruptly change signal values but that do so gradually. Finally, fuzzy attacks are usually meant to test a system for unexpected behavior, modify the values of the payload randomly, and are usually relatively simple to detect. 
All generated attacks are signal-aware and generated using an enhanced version of the \ac{READ} heuristic~\cite{MarchettiStabili}, modifying only the targeted field within each packet. 
Each attack is independently generated per \ac{CAN} ID and spaced at least one minute apart. Following CANova~\cite{nichelinipozzoli}, we select 12 specific IDs for evaluation. To prevent overfitting, training and testing sequences are drawn from disjoint portions of the dataset. In black-box settings, oracles and IDSs are trained on separate sequences to simulate differing attacker and defender environments.

\subsubsection{Car Hacking Dataset}
\label{ssec:chdata}

The publicly available \ac{CH} dataset~\cite{seo2018gids} complements ReCAN by offering real-world adversarial scenarios collected via the \ac{OBD} port during live vehicle operation. It includes five experiments: \ac{DoS}, fuzzy, RPM spoofing, gear spoofing, and an attack-free trace. We focus on the RPM and gear spoofing attacks (IDs \texttt{0316} and \texttt{043f}) as they involve \emph{payload-level manipulations} and are conducted in a realistic setting.

Unlike synthetic scenarios, these attacks were executed live with safety in mind. As such, they are inherently less aggressive, need to provide predictable outcomes (e.g., only specific IDs are attacked) and must maintain vehicular safety during execution. Nevertheless, they provide valuable insights into how adversarial perturbations behave in practical contexts.

We exclude the DoS and fuzzy attacks from our evaluation, as they are trivially detectable~\cite{refat2022detecting,seo2018gids} and not aligned with the stealth-focused objective of our adversarial approach. Due to known inconsistencies between the ``attack-free'' and ``injected'' logs~\cite{verma2022addressing}, we train all models using the untampered portions of the attack sessions themselves to avoid distribution shift.

\subsection{Selected intrusion detection systems}
Given that attacks that alter the frequency are easy to detect, an adversarial attacker will implement masquerade attacks that do not alter frequency of the network stream. Therefore, to assess the effectiveness of evasion attacks against \acp{IDS} in the automotive field, we select six commonly used~\cite{longari2023evaluating} architectures of payload-based anomaly detector. 

    \mypar{FFNN} A one-to-one autoencoder with two fully connected layers with 16 units each; this architecture is blind to replay attacks as it considers a single packet at a time, however, it is a simple solution to detect more obvious payload tampering and is included to provide a baseline reference.
    
    \mypar{CANdito~\cite{longari2022candito}} A window-to-window symmetrical autoencoder with two fully connected (128 units each) and two \ac{LSTM} layers (64 cells each), it is the most complex among the chosen networks.

    \mypar{\ac{LSTM} predictors~\cite{taylor2016anomaly}} Two window-to-one predictors, we implement two variants, a short with just two \ac{LSTM} layers having 32 cells each, and a long with four \ac{LSTM} layers, with 64 and 16 cells.

    \mypar{\acs{GRU}-based predictors} Two window-to-one predictors analogous to the short and long \ac{LSTM} variants, employing the more lightweight \acp{GRU} instead.

A final sigmoid-activated dense layer with one unit per input feature follows each individual architecture to provide an output with the correct dimensionality. While the predictor models produce an anomaly score for one packet at a time, with a rolling input window, the CANdito autoencoder reconstructs the whole window, operating with non-overlapping input sequences. We choose a window size of 40 (or 39 plus one predicted frame for the predictor models).

\subsection{Experiment 1: Adversarial attack performance evaluation}

\subsubsection{Baselines Performance}
\label{ssec:baseline_perf}

\begin{table}
\centering
\caption{Average Recall and Precision for each non-evasive attack on the ReCAN and CarHacking Dataset.}
\label{tab:baselines}
\vspace{-7px}
\resizebox{\textwidth}{!}{
\small
\begin{tabular}{|c|c|c|c|c|c|c|c|c|}
\hline
\multirow{11}{*}{\rotatebox[origin=c]{90}{\textbf{ReCAN}}} & \textbf{Attack} & \textbf{Metrics} & \textbf{FFNN} & \textbf{CANdito} & \textbf{shortLSTM} & \textbf{longLSTM} & \textbf{shortGRU} & \textbf{longGRU} \\
\cline{2-9}
& \multirow{2}{*}{Fuzzy} & recall & 0.9192 & 0.9777 & 0.8928 & 0.8430 & 0.8976 & 0.9058 \\
&  & precision & 0.7032 & 0.7758 & 0.4542 & 0.4264 & 0.5241 & 0.4763 \\
\cline{2-9}
& \multirow{2}{*}{Continuous Change} & recall & 0.4317 & 0.7758 & 0.6321 & 0.6256 & 0.6224 & 0.6190 \\
&  & precision & 0.5424 & 0.7270 & 0.2874 & 0.3294 & 0.3478 & 0.3125 \\
\cline{2-9}
& \multirow{2}{*}{Change to Minimum} & recall & 0.5158 & 0.7845 & 0.5307 & 0.5550 & 0.5224 & 0.5116 \\
&  & precision & 0.7152 & 0.9184 & 0.2852 & 0.2916 & 0.3761 & 0.2839 \\
\cline{2-9}
& \multirow{2}{*}{Injection Replay} & recall & 0.0000 & 0.4215 & 0.6640 & 0.5447 & 0.6927 & 0.6893 \\
&  & precision & 0.0000 & 0.6513 & 0.2436 & 0.3268 & 0.2718 & 0.2494 \\
\cline{2-9}
& \multirow{2}{*}{Masquerade Replay} & recall & 0.0006 & 0.3604 & 0.2743 & 0.2557 & 0.2070 & 0.2214 \\
&  & precision & 0.0067 & 0.5024 & 0.1522 & 0.1728 & 0.1690 & 0.1366 \\
\hline
\hline
\multirow{4}{*}{\rotatebox[origin=c]{90}{\textbf{CarHack.}}} 
& \multirow{2}{*}{RPM Spoofing} & recall & 1.0000 & 1.0000 & 1.0000 & 1.0000 & 1.0000 & 1.0000 \\
&  & precision & 1.0000 & 1.0000 & 0.9066 & 0.9066 & 0.9066 & 0.9066 \\
\cline{2-9}
& \multirow{2}{*}{Gear Spoofing} & recall & 1.0000 & 1.0000 & 1.0000 & 1.0000 & 1.0000 & 1.0000 \\
&  & precision & 0.9950 & 0.9655 & 0.9090 & 0.9090 & 0.9090 & 0.9090 \\
\hline
\end{tabular}
}
\vspace{-10px}
\end{table}

Table~\ref{tab:baselines} reports the average recall and precision for various non-evasive attacks evaluated across two datasets—ReCAN and CarHacking—using a range of IDS models: FFNN, CANdito, shortLSTM, longLSTM, shortGRU, and longGRU. These metrics provide a baseline understanding of each model’s ability to detect attacks under non-evasive conditions.

On the ReCAN dataset, results align with expectations. The \textit{Fuzzy} attack, which significantly disrupts data patterns, yields high recall across all models, with CANdito showing the best performance. However, precision drops notably—especially in shortLSTM—indicating non null false positive rates (note that this is not particularly relevant for our experiments, since the adversarial samples do not affect non-tampered data). Attacks such as \textit{Continuous Change} and \textit{Change to Minimum} show moderate recall, particularly in CANdito and longLSTM, but again suffer from low precision, especially in shortLSTM and shortGRU. The \textit{Injection Replay} and \textit{Masquerade Replay} attacks are particularly challenging to detect, showing very low recall and precision across all models. For instance, FFNN fails almost entirely to detect Injection Replay, due to its lack of temporal context, while sliding-window models perform slightly better, benefiting from sequential input awareness.

In contrast, the CarHacking dataset yields significantly stronger results. Both the \textit{RPM Spoofing} and \textit{Gear Spoofing} attacks are consistently detected, with all models achieving perfect recall. Precision is similarly high across the board, with CANdito slightly outperforming others on the \textit{Gear Spoofing} attack. These results highlight that real-world injected attacks, potentially due to the safety limitations in generating the dataset, are more easily detected than synthetic ones, and the similarity in performance across the RPM and Gear spoofing attacks is expected, as both rely on the same injection-based attack strategy.

\subsubsection{White-Box scenario}
\label{ssec:wb_scenario}

\begin{table}\centering
\caption{Results on the ReCAN dataset in the white-box, grey-box (first part of the table, the white diagonal line represent the white-box), and black-box scenarios for the fuzzy attack.}
\vspace{-7px}
\label{tab:wb_recan_fuzzy}
\resizebox{\textwidth}{!}{
\begin{tabular}{|c|c|c|cccccc|}
\hline \multicolumn{3}{|c|}{White-box and Grey-box} & \multicolumn{6}{c|}{\textbf{TPR ($\Delta$)}} \\ \hline
\textbf{Oracle} & \textbf{Algorithm} & \textbf{AP} & \textbf{FFNN} & \textbf{CANdito} & \textbf{Short LSTM} & \textbf{Long LSTM} & \textbf{Short GRU} & \textbf{Long GRU} \\ \hline
\multirow{3}{*}{FFNN} & Decay BIM & 0.02 & 0.81 ({\color[rgb]{1,0.10311111111110893,0}-11.21\%}) & \cellcolor{lightgray!30} 0.98 ({\color[rgb]{0,0,0}0.0\%}) & \cellcolor{lightgray!30} 0.89 ({\color[rgb]{1,0.6,0}-0.12\%}) & \cellcolor{lightgray!30} 0.84 ({\color[rgb]{0,0.0033333333333329662,0}0.07\%}) & \cellcolor{lightgray!30} 0.9 ({\color[rgb]{1,0.6,0}-0.06\%}) & \cellcolor{lightgray!30} 0.9 ({\color[rgb]{1,0.6,0}-0.29\%}) \\
& L2 BIM & 0.18 & 0.43 ({\color[rgb]{1,0,0}-48.99\%}) & \cellcolor{lightgray!30} 0.98 ({\color[rgb]{0,0,0}0.0\%}) & \cellcolor{lightgray!30} 0.89 ({\color[rgb]{1,0.6,0}-0.71\%}) & \cellcolor{lightgray!30} 0.83 ({\color[rgb]{1,0.6,0}-1.29\%}) & \cellcolor{lightgray!30} 0.89 ({\color[rgb]{1,0.6,0}-0.68\%}) & \cellcolor{lightgray!30} 0.9 ({\color[rgb]{1,0.6,0}-0.71\%}) \\
& DeepFool & 0.19 & 0.58 ({\color[rgb]{1,0,0}-33.67\%}) & \cellcolor{lightgray!30} 0.98 ({\color[rgb]{0,0,0}0.0\%}) & \cellcolor{lightgray!30} 0.88 ({\color[rgb]{1,0.6,0}-0.94\%}) & \cellcolor{lightgray!30} 0.82 ({\color[rgb]{1,0.6,0}-1.94\%}) & \cellcolor{lightgray!30} 0.89 ({\color[rgb]{1,0.6,0}-0.8\%}) & \cellcolor{lightgray!30} 0.89 ({\color[rgb]{1,0.6,0}-1.26\%}) \\
\hline
\multirow{3}{*}{CANdito} & Decay BIM & 0.0 & \cellcolor{lightgray!30} 0.92 ({\color[rgb]{0,0,0}0.0\%}) & 0.98 ({\color[rgb]{0,0,0}0.0\%}) & \cellcolor{lightgray!30} 0.89 ({\color[rgb]{0,0,0}0.0\%}) & \cellcolor{lightgray!30} 0.84 ({\color[rgb]{0,0,0}0.0\%}) & \cellcolor{lightgray!30} 0.9 ({\color[rgb]{0,0,0}0.0\%}) & \cellcolor{lightgray!30} 0.91 ({\color[rgb]{0,0,0}0.0\%}) \\
\hline

\end{tabular}
}
\vspace{-0.3cm}

\end{table}

\begin{table}\centering
\caption{Results on the ReCAN dataset in the white-box, grey-box (first part of the table, the white diagonal line represent the white-box), and black-box scenarios for the continuous change attack.}
\vspace{-7px}
\label{tab:wb_recan_continuous}
\resizebox{\textwidth}{!}{
\begin{tabular}{|c|c|c|cccccc|}
\hline\multicolumn{3}{|c|}{White-box and Grey-box} & \multicolumn{6}{c|}{\textbf{TPR ($\Delta$)}} \\ \hline
\textbf{Oracle} & \textbf{Algorithm} & \textbf{AP} & \textbf{FFNN} & \textbf{CANdito} & \textbf{Short LSTM} & \textbf{Long LSTM} & \textbf{Short GRU} & \textbf{Long GRU} \\ \hline
\multirow{3}{*}{FFNN} & Decay BIM & 0.02 & 0.2 ({\color[rgb]{1,0,0}-23.64\%}) & \cellcolor{lightgray!30} 0.78 ({\color[rgb]{0,0,0}0.0\%}) & \cellcolor{lightgray!30} 0.62 ({\color[rgb]{1,0.6,0}-0.76\%}) & \cellcolor{lightgray!30} 0.62 ({\color[rgb]{1,0.6,0}-0.63\%}) & \cellcolor{lightgray!30} 0.61 ({\color[rgb]{1,0.6,0}-0.99\%}) & \cellcolor{lightgray!30} 0.62 ({\color[rgb]{1,0.6,0}-0.18\%}) \\
& L2 BIM & 0.06 & 0.07 ({\color[rgb]{1,0,0}-35.68\%}) & \cellcolor{lightgray!30} 0.78 ({\color[rgb]{0,0,0}0.0\%}) & \cellcolor{lightgray!30} 0.6 ({\color[rgb]{1,0.6,0}-3.41\%}) & \cellcolor{lightgray!30} 0.58 ({\color[rgb]{1,0.6,0}-4.33\%}) & \cellcolor{lightgray!30} 0.59 ({\color[rgb]{1,0.6,0}-2.86\%}) & \cellcolor{lightgray!30} 0.6 ({\color[rgb]{1,0.6,0}-2.37\%}) \\
& DeepFool & 0.04 & 0.21 ({\color[rgb]{1,0,0}-22.11\%}) & \cellcolor{lightgray!30} 0.78 ({\color[rgb]{0,0,0}0.0\%}) & \cellcolor{lightgray!30} 0.62 ({\color[rgb]{1,0.6,0}-0.86\%}) & \cellcolor{lightgray!30} 0.61 ({\color[rgb]{1,0.6,0}-1.68\%}) & \cellcolor{lightgray!30} 0.62 ({\color[rgb]{1,0.6,0}-0.1\%}) & \cellcolor{lightgray!30} 0.62 ({\color[rgb]{0,0.008888888888889945,0}0.18\%}) \\
\hline
\multirow{3}{*}{CANdito} & Decay BIM & 0.0 & \cellcolor{lightgray!30} 0.43 ({\color[rgb]{0,0,0}0.0\%}) & 0.78 ({\color[rgb]{0,0,0}0.0\%}) & \cellcolor{lightgray!30} 0.63 ({\color[rgb]{0,0,0}0.0\%}) & \cellcolor{lightgray!30} 0.63 ({\color[rgb]{0,0,0}0.0\%}) & \cellcolor{lightgray!30} 0.62 ({\color[rgb]{0,0,0}0.0\%}) & \cellcolor{lightgray!30} 0.62 ({\color[rgb]{0,0,0}0.0\%}) \\
\hline

\end{tabular}
}
\vspace{-0.3cm}

\end{table}

\begin{table}\centering
\caption{Results on the ReCAN dataset in the white-box, grey-box (first part of the table, the white diagonal line represent the white-box), and black-box scenarios for the change to minimum attack.}
\vspace{-7px}
\label{tab:wb_recan_minimum}
\resizebox{\textwidth}{!}{
\begin{tabular}{|c|c|c|cccccc|}
\hline \multicolumn{3}{|c|}{White-box and Grey-box} & \multicolumn{6}{c|}{\textbf{TPR ($\Delta$)}} \\ \hline
\textbf{Oracle} & \textbf{Algorithm} & \textbf{AP} & \textbf{FFNN} & \textbf{CANdito} & \textbf{Short LSTM} & \textbf{Long LSTM} & \textbf{Short GRU} & \textbf{Long GRU} \\ \hline
\multirow{3}{*}{FFNN} & Decay BIM & 0.02 & 0.31 ({\color[rgb]{1,0,0}-20.7\%}) & \cellcolor{lightgray!30} 0.85 ({\color[rgb]{0,0.32454004329004416,0}6.49\%}) & \cellcolor{lightgray!30} 0.65 ({\color[rgb]{0,0.6216666666666665,0}12.43\%}) & \cellcolor{lightgray!30} 0.68 ({\color[rgb]{0,0.6045726495726494,0}12.09\%}) & \cellcolor{lightgray!30} 0.68 ({\color[rgb]{0,0.7648717948717948,0}15.3\%}) & \cellcolor{lightgray!30} 0.67 ({\color[rgb]{0,0.8014957264957273,0}16.03\%}) \\
& L2 BIM & 0.09 & 0.08 ({\color[rgb]{1,0,0}-43.68\%}) & \cellcolor{lightgray!30} 0.81 ({\color[rgb]{0,0.14398448773448813,0}2.88\%}) & \cellcolor{lightgray!30} 0.61 ({\color[rgb]{0,0.4066666666666663,0}8.13\%}) & \cellcolor{lightgray!30} 0.61 ({\color[rgb]{0,0.2779059829059827,0}5.56\%}) & \cellcolor{lightgray!30} 0.61 ({\color[rgb]{0,0.46209401709401743,0}9.24\%}) & \cellcolor{lightgray!30} 0.63 ({\color[rgb]{0,0.5698290598290595,0}11.4\%}) \\
& DeepFool & 0.06 & 0.27 ({\color[rgb]{1,0,0}-24.89\%}) & \cellcolor{lightgray!30} 0.85 ({\color[rgb]{0,0.3401124338624345,0}6.8\%}) & \cellcolor{lightgray!30} 0.63 ({\color[rgb]{0,0.5133333333333329,0}10.27\%}) & \cellcolor{lightgray!30} 0.64 ({\color[rgb]{0,0.434017094017094,0}8.68\%}) & \cellcolor{lightgray!30} 0.64 ({\color[rgb]{0,0.6132051282051282,0}12.26\%}) & \cellcolor{lightgray!30} 0.64 ({\color[rgb]{0,0.6437179487179484,0}12.87\%}) \\
\hline
\end{tabular}
}
\vspace{-0.3cm}

\end{table}

\begin{table}\centering
\caption{Results on the ReCAN dataset in the white-box, grey-box (first part of the table, the white diagonal line represent the white-box), and black-box scenarios for the masquerade replay attack.}
\vspace{-7px}
\label{tab:wb_recan_fullreplay}
\resizebox{\textwidth}{!}{
\begin{tabular}{|c|c|c|cccccc|}
\hline\multicolumn{3}{|c|}{White-box and Grey-box} & \multicolumn{6}{c|}{\textbf{TPR ($\Delta$)}} \\ \hline
\textbf{Oracle} & \textbf{Algorithm} & \textbf{AP} & \textbf{FFNN} & \textbf{CANdito} & \textbf{Short LSTM} & \textbf{Long LSTM} & \textbf{Short GRU} & \textbf{Long GRU} \\ \hline
\multirow{3}{*}{FFNN} & Decay BIM & 0.0 & 0.0 ({\color[rgb]{1,0.6,0}-0.06\%}) & \cellcolor{lightgray!30} 0.36 ({\color[rgb]{0,0,0}0.0\%}) & \cellcolor{lightgray!30} 0.27 ({\color[rgb]{0,0,0}0.0\%}) & \cellcolor{lightgray!30} 0.26 ({\color[rgb]{0,0,0}0.0\%}) & \cellcolor{lightgray!30} 0.21 ({\color[rgb]{0,0,0}0.0\%}) & \cellcolor{lightgray!30} 0.22 ({\color[rgb]{0,0,0}0.0\%}) \\
& L2 BIM & 0.0 & 0.0 ({\color[rgb]{1,0.6,0}-0.06\%}) & \cellcolor{lightgray!30} 0.36 ({\color[rgb]{0,0,0}0.0\%}) & \cellcolor{lightgray!30} 0.27 ({\color[rgb]{0,0,0}0.0\%}) & \cellcolor{lightgray!30} 0.26 ({\color[rgb]{0,0,0}0.0\%}) & \cellcolor{lightgray!30} 0.21 ({\color[rgb]{0,0,0}0.0\%}) & \cellcolor{lightgray!30} 0.22 ({\color[rgb]{0,0,0}0.0\%}) \\
& DeepFool & 0.0 & 0.0 ({\color[rgb]{1,0.6,0}-0.06\%}) & \cellcolor{lightgray!30} 0.36 ({\color[rgb]{0,0,0}0.0\%}) & \cellcolor{lightgray!30} 0.27 ({\color[rgb]{0,0,0}0.0\%}) & \cellcolor{lightgray!30} 0.26 ({\color[rgb]{0,0,0}0.0\%}) & \cellcolor{lightgray!30} 0.21 ({\color[rgb]{0,0,0}0.0\%}) & \cellcolor{lightgray!30} 0.22 ({\color[rgb]{0,0,0}0.0\%}) \\
\hline
\multirow{3}{*}{CANdito} & Decay BIM & 0.01 & \cellcolor{lightgray!30} 0.01 ({\color[rgb]{0,0.02777777777777775,0}0.56\%}) & 0.24 ({\color[rgb]{1,0.007407407407407307,0}-12.41\%}) & \cellcolor{lightgray!30} 0.28 ({\color[rgb]{0,0.0477777777777777,0}0.96\%}) & \cellcolor{lightgray!30} 0.27 ({\color[rgb]{0,0.05444444444444446,0}1.09\%}) & \cellcolor{lightgray!30} 0.22 ({\color[rgb]{0,0.06055555555555564,0}1.21\%}) & \cellcolor{lightgray!30} 0.23 ({\color[rgb]{0,0.026666666666666505,0}0.53\%}) \\
\hline

\end{tabular}
}
\vspace{-0.3cm}

\end{table}

\begin{table}\centering
\caption{Results on the ReCAN dataset in the white-box, grey-box (first part of the table, the white diagonal line represent the white-box), and black-box scenarios for the injection replay attack.}
\vspace{-7px}
\label{tab:wb_recan_injection}
\resizebox{\textwidth}{!}{
\begin{tabular}{|c|c|c|cccccc|}
\hline \multicolumn{3}{|c|}{White-box and Grey-box} & \multicolumn{6}{c|}{\textbf{TPR ($\Delta$)}} \\ \hline
\textbf{Oracle} & \textbf{Algorithm} & \textbf{AP} & \textbf{FFNN} & \textbf{CANdito} & \textbf{Short LSTM} & \textbf{Long LSTM} & \textbf{Short GRU} & \textbf{Long GRU} \\ \hline
\multirow{3}{*}{FFNN} & Decay BIM & 0.0 & 0.0 ({\color[rgb]{0,0,0}0.0\%}) & \cellcolor{lightgray!30} 0.42 ({\color[rgb]{0,0,0}0.0\%}) & \cellcolor{lightgray!30} 0.66 ({\color[rgb]{0,0,0}0.0\%}) & \cellcolor{lightgray!30} 0.54 ({\color[rgb]{0,0,0}0.0\%}) & \cellcolor{lightgray!30} 0.69 ({\color[rgb]{0,0,0}0.0\%}) & \cellcolor{lightgray!30} 0.69 ({\color[rgb]{0,0,0}0.0\%}) \\
& L2 BIM & 0.0 & 0.0 ({\color[rgb]{0,0,0}0.0\%}) & \cellcolor{lightgray!30} 0.42 ({\color[rgb]{0,0,0}0.0\%}) & \cellcolor{lightgray!30} 0.66 ({\color[rgb]{0,0,0}0.0\%}) & \cellcolor{lightgray!30} 0.54 ({\color[rgb]{0,0,0}0.0\%}) & \cellcolor{lightgray!30} 0.69 ({\color[rgb]{0,0,0}0.0\%}) & \cellcolor{lightgray!30} 0.69 ({\color[rgb]{0,0,0}0.0\%}) \\
& DeepFool & 0.0 & 0.0 ({\color[rgb]{0,0,0}0.0\%}) & \cellcolor{lightgray!30} 0.42 ({\color[rgb]{0,0,0}0.0\%}) & \cellcolor{lightgray!30} 0.66 ({\color[rgb]{0,0,0}0.0\%}) & \cellcolor{lightgray!30} 0.54 ({\color[rgb]{0,0,0}0.0\%}) & \cellcolor{lightgray!30} 0.69 ({\color[rgb]{0,0,0}0.0\%}) & \cellcolor{lightgray!30} 0.69 ({\color[rgb]{0,0,0}0.0\%}) \\
\hline
\multirow{3}{*}{CANdito} & Decay BIM & 0.0 & \cellcolor{lightgray!30} 0.0 ({\color[rgb]{0,0,0}0.0\%}) & 0.41 ({\color[rgb]{1,0.6,0}-1.01\%}) & \cellcolor{lightgray!30} 0.66 ({\color[rgb]{0,0,0}0.0\%}) & \cellcolor{lightgray!30} 0.54 ({\color[rgb]{0,0,0}0.0\%}) & \cellcolor{lightgray!30} 0.69 ({\color[rgb]{0,0,0}0.0\%}) & \cellcolor{lightgray!30} 0.69 ({\color[rgb]{0,0,0}0.0\%}) \\
\hline

\end{tabular}
}
\vspace{-0.3cm}

\end{table}

\begin{table}\centering
\caption{Results on the CarHacking dataset in the white-box, grey-box (first part of the table, the white diagonal line represent the white-box), and black-box scenarios for the RPM spoofing attack.}
\label{tab:carhacking_rpm}
\vspace{-7px}
\resizebox{\textwidth}{!}{
\begin{tabular}{|c|c|c|cccccc|}

\hline\multicolumn{3}{|c|}{White-box and Grey-box} & \multicolumn{6}{c|}{\textbf{TPR ($\Delta$)}} \\ \hline
\textbf{Oracle} & \textbf{Algorithm} & \textbf{AP} & \textbf{FFNN} & \textbf{CANdito} & \textbf{Short LSTM} & \textbf{Long LSTM} & \textbf{Short GRU} & \textbf{Long GRU} \\ \hline
\multirow{3}{*}{FFNN} & Decay BIM & 0.0 & 1.0 ({\color[rgb]{0,0,0}0.0\%}) & \cellcolor{lightgray!30} 1.0 ({\color[rgb]{0,0,0}0.0\%}) & \cellcolor{lightgray!30} 1.0 ({\color[rgb]{0,0,0}0.0\%}) & \cellcolor{lightgray!30} 1.0 ({\color[rgb]{0,0,0}0.0\%}) & \cellcolor{lightgray!30} 1.0 ({\color[rgb]{0,0,0}0.0\%}) & \cellcolor{lightgray!30} 1.0 ({\color[rgb]{0,0,0}0.0\%}) \\
& L2 BIM & 0.68 & 1.0 ({\color[rgb]{0,0,0}0.0\%}) & \cellcolor{lightgray!30} 1.0 ({\color[rgb]{0,0,0}0.0\%}) & \cellcolor{lightgray!30} 1.0 ({\color[rgb]{0,0,0}0.0\%}) & \cellcolor{lightgray!30} 1.0 ({\color[rgb]{0,0,0}0.0\%}) & \cellcolor{lightgray!30} 1.0 ({\color[rgb]{0,0,0}0.0\%}) & \cellcolor{lightgray!30} 1.0 ({\color[rgb]{0,0,0}0.0\%}) \\
& DeepFool & 0.0 & 0.15 ({\color[rgb]{1,0,0}-85.02\%}) & \cellcolor{lightgray!30} 1.0 ({\color[rgb]{0,0,0}0.0\%}) & \cellcolor{lightgray!30} 0.15 ({\color[rgb]{1,0,0}-84.51\%}) & \cellcolor{lightgray!30} 0.15 ({\color[rgb]{1,0,0}-84.51\%}) & \cellcolor{lightgray!30} 0.15 ({\color[rgb]{1,0,0}-84.51\%}) & \cellcolor{lightgray!30} 0.15 ({\color[rgb]{1,0,0}-84.51\%}) \\
\hline
\multirow{3}{*}{CANdito} & Decay BIM & 0.0 & \cellcolor{lightgray!30} 1.0 ({\color[rgb]{0,0,0}0.0\%}) & 1.0 ({\color[rgb]{0,0,0}0.0\%}) & \cellcolor{lightgray!30} 1.0 ({\color[rgb]{0,0,0}0.0\%}) & \cellcolor{lightgray!30} 1.0 ({\color[rgb]{0,0,0}0.0\%}) & \cellcolor{lightgray!30} 1.0 ({\color[rgb]{0,0,0}0.0\%}) & \cellcolor{lightgray!30} 1.0 ({\color[rgb]{0,0,0}0.0\%}) \\
\hline

\end{tabular}
}
\vspace{-0.3cm}

\end{table}
\begin{table}\centering
\caption{Results on the CarHacking dataset in the white-box, grey-box (first part of the table, the white diagonal line represent the white-box), and black-box scenarios for the gear spoofing attack.}
\label{tab:carhacking_gear}
\vspace{-7px}
\resizebox{\textwidth}{!}{
\begin{tabular}{|c|c|c|cccccc|}
\hline\multicolumn{3}{|c|}{White-box and Grey-box} & \multicolumn{6}{c|}{\textbf{TPR ($\Delta$)}} \\ \hline
\textbf{Oracle} & \textbf{Algorithm} & \textbf{AP} & \textbf{FFNN} & \textbf{CANdito} & \textbf{Short LSTM} & \textbf{Long LSTM} & \textbf{Short GRU} & \textbf{Long GRU} \\ \hline
\multirow{3}{*}{FFNN} & Decay BIM & 0.0 & 1.0 ({\color[rgb]{0,0,0}0.0\%}) & \cellcolor{lightgray!30} 1.0 ({\color[rgb]{0,0,0}0.0\%}) & \cellcolor{lightgray!30} 1.0 ({\color[rgb]{0,0,0}0.0\%}) & \cellcolor{lightgray!30} 1.0 ({\color[rgb]{0,0,0}0.0\%}) & \cellcolor{lightgray!30} 1.0 ({\color[rgb]{0,0,0}0.0\%}) & \cellcolor{lightgray!30} 1.0 ({\color[rgb]{0,0,0}0.0\%}) \\
& L2 BIM & 0.0 & 1.0 ({\color[rgb]{0,0,0}0.0\%}) & \cellcolor{lightgray!30} 1.0 ({\color[rgb]{0,0,0}0.0\%}) & \cellcolor{lightgray!30} 1.0 ({\color[rgb]{0,0,0}0.0\%}) & \cellcolor{lightgray!30} 1.0 ({\color[rgb]{0,0,0}0.0\%}) & \cellcolor{lightgray!30} 1.0 ({\color[rgb]{0,0,0}0.0\%}) & \cellcolor{lightgray!30} 1.0 ({\color[rgb]{0,0,0}0.0\%}) \\
& DeepFool & 0.0 & 0.17 ({\color[rgb]{1,0,0}-83.46\%}) & \cellcolor{lightgray!30} 0.14 ({\color[rgb]{1,0,0}-86.0\%}) & \cellcolor{lightgray!30} 0.17 ({\color[rgb]{1,0,0}-82.9\%}) & \cellcolor{lightgray!30} 0.17 ({\color[rgb]{1,0,0}-82.9\%}) & \cellcolor{lightgray!30} 0.17 ({\color[rgb]{1,0,0}-82.9\%}) & \cellcolor{lightgray!30} 0.17 ({\color[rgb]{1,0,0}-82.9\%}) \\
\hline
\multirow{3}{*}{CANdito} & Decay BIM & 0.0 & \cellcolor{lightgray!30} 1.0 ({\color[rgb]{0,0,0}0.0\%}) & 1.0 ({\color[rgb]{0,0,0}0.0\%}) & \cellcolor{lightgray!30} 1.0 ({\color[rgb]{0,0,0}0.0\%}) & \cellcolor{lightgray!30} 1.0 ({\color[rgb]{0,0,0}0.0\%}) & \cellcolor{lightgray!30} 1.0 ({\color[rgb]{0,0,0}0.0\%}) & \cellcolor{lightgray!30} 1.0 ({\color[rgb]{0,0,0}0.0\%}) \\
\hline

\end{tabular}
}
\vspace{-0.3cm}

\end{table}

In this experiment, we compare the performances of all six \acp{IDS} over all the available attacks in the white-box scenario. Tables,~\ref{tab:wb_recan_fuzzy},~\ref{tab:wb_recan_continuous},~\ref{tab:wb_recan_minimum},~\ref{tab:wb_recan_fullreplay}, and~\ref{tab:wb_recan_injection} show the aggregated results for the adversarial attacks on the ReCAN dataset, while Tables~\ref{tab:carhacking_rpm} and~\ref{tab:carhacking_gear} present the results on the CarHacking dataset. Note that the performance metric values in each table are averaged over all evaluated IDs. Since we test the evasive packets generated by each oracle model against all the other available architectures on the same dataset, the diagonal in these tables represents the results of the white-box scenario, while the remaining part shows the results of the grey-box one.

\mypar{ReCAN} Overall, the performances on the ReCAN datasets in the white-box scenario demonstrate the effectiveness of the evasion process. Across all the attacks, multiple algorithms manage to lower the detection rate of the respective detection systems by 10\% to 60\%. As presented in Table~\ref{tab:evasivealgorithmsperformances}, the oracles in the white-box scenario significantly outperform the others in lowering the detection rate.

Among all attacks, the fuzzy attack (see Table~\ref{tab:wb_recan_fuzzy}) is predictably the easiest to detect due to the random fluctuations in the values of the tampered signals. The intent behind the fuzzing attack is often to investigate the behavior of a system or study its interaction with the \acp{IDS}, rather than to produce a specific effect. Notably, it is also the attack with the highest white-box evasion performance, with DeepFool repeatedly causing a detection rate loss exceeding 50\% compared to the baseline. Interestingly, CANdito maintains a significantly lower performance loss than the other algorithms.

The continuous change and change-to-minimum attacks (see Tables~\ref{tab:wb_recan_continuous} and ~\ref{tab:wb_recan_minimum}) exhibit comparable baseline performances, with the former being slightly easier to detect (with a difference of about 5-10\% in TPR), likely due to the randomly generated target value potentially falling outside the normal behavior of the signal. As expected, for similar reasons, it is also easier to find an evasive sample for the continuous change attack. In these attacks, the intruder replays previous packets while tampering with just one signal field— with a minimum length of 9 bits— making the payload progressively easier to detect as the attack continues. Unfortunately, this behavior also causes the algorithms to strongly perturb the target signal, often resulting in an example that closely resembles a packet from the attack-free scenario, meaning that while the evasion is successful, the intended effect is lost. Notably, while DeepFool performs best in the fuzzy attack, in these cases it is the L2 BIM algorithm that consistently achieves a 40\% to 50

The masquerade replay attack (see Table~\ref{tab:wb_recan_fullreplay}) justifiably has the lowest detection rates in the baselines, due to the attack signals being valid sequences in themselves, simply inserted at a different moment on the network by the attacker. Therefore, the system can only leverage the semantic discontinuity in the flow of messages for its classification. The FFNN algorithm is rendered completely useless, meaning no evasion is necessary or achieved. CANdito with the L2 BIM algorithm performs best in terms of detection loss, with a 23\% reduction, but overall the evasion capabilities are relatively low, likely due to the already low detection performances of the detection systems.

Finally, in the injection replay scenario (see Table~\ref{tab:wb_recan_injection}), the attacker interleaves additional replayed packets into the normal traffic so that the content is analogous to full replay sequences, but discontinuities are present at each injection. This is the only attack against which the predictors outperform CANdito in the baseline test. Similarly, the drop in detection rate is more pronounced in the predictor models than in CANdito. As in the previous case, since it does not consider temporal windows for detection, the FFNN is rendered completely useless.

\mypar{CarHacking} Interestingly, the results on the dataset are heavily dependent on the evasion algorithm used. In fact, the DeepFool algorithm is the only one to achieve any kind of evasion in both the gear and RPM spoofing attack scenarios (see Tables~\ref{tab:carhacking_gear}
and~\ref{tab:carhacking_rpm}). In these attacks, however, the drop in detection performance is up to 85\% for the FFNN and remains consistently high for all but CANdito, with the only exception being the Long GRU model in the RPM spoofing attack.

\subsubsection{Grey-Box scenario}
\label{ssec:gb_scenario}

In this experiment, we compare the performances of all six \acp{IDS} over all the available attacks in the grey-box scenario. Tables,~\ref{tab:wb_recan_fuzzy},~\ref{tab:wb_recan_continuous},~\ref{tab:wb_recan_minimum},~\ref{tab:wb_recan_fullreplay}, and~\ref{tab:wb_recan_injection} show the aggregated results for the adversarial attacks on the ReCAN dataset, while Tables~\ref{tab:carhacking_rpm} and~\ref{tab:carhacking_gear} on the CarHacking dataset. We remember the reader that since we test the evasive packets generated by each oracle model against all the other available architectures on the same dataset, the diagonal in these tables represents the results of the white-box scenario, while the remaining part the results of the grey-box one. 

\mypar{ReCAN} We refrain from evaluating each attack scenario again due to repetitiveness and space constraints and instead highlight the most notable characteristics of the grey-box performances. Overall, the results are, as expected, worse than in the white-box scenario and better than in the black-box one. In some attack scenarios where the white-box attacks are effective, such as the fuzzy and change-to-minimum attacks, the performance of using oracles different from the detection model, although trained on the same data, drops significantly. The delta in performance loss is less prominent in the already least-performing attacks.
Probably the most notable characteristic of the grey-box scenario is that the autoencoder models, namely CANdito and the FFNN, when used to attack the predictive models, perform poorly enough to even increase the detection performance, as is the case with the change-to-minimum attack. On the other hand, the FFNN used as a defense system seems to increase its detection rate in grey-box attacks in the masquerade and injection replay scenarios, mostly because these perturbations sometimes alter the shape of the packet from one identical to previous traffic to one different enough to be detected. Nevertheless, the overall detection performances of the algorithm in such scenarios still render it ineffective. 

\mypar{CarHacking}
Aside from the lack of effectiveness of the BIM algorithms, the grey-box scenarios for the CarHacking dataset mostly confirm what was already observed in the white-box scenario. With DeepFool, the FFNN is extremely effective against most models, while the other oracles lose some effectiveness but still maintain evasion capabilities. Interestingly, the FFNN also proves to be effective as an intrusion detection system, apparently contradicting the results from the ReCAN dataset. The hypothesis for why this occurs is that the ease of detection for the FFNN (and partially for all detection systems) stems from the simplicity of the signal characteristics in the CarHacking dataset, which are almost always constant.
Notably, CANdito is the only IDS that is not evaded by any oracle, but it is also the only oracle that does not find any suitable evasive point.

\subsubsection{Black-Box scenario}
\label{ssec:bb_scenario}
Finally, we compare the performances of all six \acp{IDS} over all the available attacks in the black-box scenario. Tables,~\ref{tab:wb_recan_fuzzy},~\ref{tab:wb_recan_continuous},~\ref{tab:wb_recan_minimum},~\ref{tab:wb_recan_fullreplay}, and~\ref{tab:wb_recan_injection} show the aggregated results for the adversarial attacks on the ReCAN dataset, while Tables~\ref{tab:carhacking_rpm} and~\ref{tab:carhacking_gear} on the CarHacking dataset. The second part of the table represents the black-box scenario. It is interesting to highlight that the diagonal line of the black-box tables represents the case where the oracle and IDS share the same architecture, but the models have been trained on different datasets. However, the identical architecture does not appear to provide a noticeable increase in evasion capabilities as long as the training data is different.

\mypar{ReCAN} As expected, the performances of the oracles in the black-box scenario are significantly lower than in the white-box scenario. However, the DeepFool algorithm maintains a consistent 20\% drop in detection rate for the predictive models in all attacks except the masquerade injection case. In the black-box scenario, the predictive models remain the overall better oracles, consistently achieving better evasion performances than the autoencoders, with both the FFNN and CANdito consistently showing less than a 10\% loss in detection rate. Nonetheless, in terms of detection system performance, CANdito remains the hardest to evade, often experiencing a 5-20\% smaller drop in detection rate than the other detectors.

\mypar{CarHacking}
The results in the black-box scenario for the CarHacking dataset appear similar to those in the white- and grey-box scenarios, with the predictive oracles being able to find evasive points only against predictive IDSs, and CANdito achieving a perfect TPR against all oracles. However, the FFNN performs exceptionally well as an oracle against all but CANdito in the gear spoofing attack, while it fails completely in the RPM spoofing attack.

\subsubsection{Discussion on results}
Tables~\ref{tab:evasivealgorithmsperformances} and~\ref{tab:aggregatedresults} provide aggregated results of the performances of the various evasive algorithms, oracles, and IDSs in all scenarios and over all the tested attacks. 

It is evident from Table~\ref{tab:evasivealgorithmsperformances}  that the difference in performance between the white-box and the grey-/black-box scenarios suggests that knowing the exact model is significantly more effective than having access only to the training dataset, which appears to be closer to not having any information at all. The table also shows that DeepFool is capable of achieving much higher perturbations without being detected, making it the most effective evasion algorithm (note that while the lack of evasion capabilities of the BIM algorithms on the CarHacking dataset skews the results toward DeepFool, similar although attenuated results would be observed if only the ReCAN dataset were considered).

Table~\ref{tab:aggregatedresults} combines the results across IDSs and oracles, highlighting that while CANdito, as expected, remains the most effective intrusion detection algorithm across all scenarios, it is also the worst-performing oracle, with even a slight average increase in detection in the grey-box scenario. Overall, the predictor architectures are significantly more effective at achieving evasive samples, with the Short LSTM model performing best in all but the black-box scenario.

\begin{table}
    \centering
        \caption{Average perturbation and delta TPR of the different evasive algorithms depending on the scenario.}
        \vspace{-7px}
    \label{tab:evasivealgorithmsperformances}
    \resizebox{\textwidth}{!}{
    \begin{tabular}{|c|c|c|c|c|c|c|c|c|c|}
\hline & \multicolumn{3}{c|}{\textbf{White Box}} & \multicolumn{3}{c|}{\textbf{Grey Box}} & \multicolumn{3}{c|}{\textbf{Black Box}} \\ \cline{2-10}
& Decay BIM & L2 BIM & DeepFool & Decay BIM & L2 BIM & DeepFool & Decay BIM & L2 BIM & DeepFool \\ \hline
Avg. Perturbation & 0.02 & 0.08 & 0.17 & 0.02 & 0.08 & 0.17 & 0.01 & 0.05 & 0.15 \\ \hline
Avg. Delta Recall & -7.76\% & -20.13\% & -30.88\% & -0.82\% & -5.71\% & -10.44\% & -2.02\% & -5.21\% & -10.41\% \\ \hline
    \end{tabular}
    }
\vspace{-10px}
\end{table}

\begin{table}[h]\centering
\caption{Combined Results of IDSs and Oracles for Whitebox, Greybox, and Blackbox Scenarios.}
\vspace{-7px}
\label{tab:aggregatedresults}
\resizebox{.7\textwidth}{!}{
\begin{tabular}{|c|c|c|c|c|c|c|}
\hline
\multicolumn{7}{|c|}{\textbf{Average IDS performances over all attacks}} \\
\hline \textbf{Scenario} & \textbf{FFNN} & \textbf{CANdito} & \textbf{Short LSTM} & \textbf{Long LSTM} & \textbf{Short GRU} & \textbf{Long GRU} \\ \hline
White Box & 0.35 & 0.71 & 0.47 & 0.45 & 0.49 & 0.48 \\ \hline
Grey Box & 0.52 & 0.72 & 0.64 & 0.62 & 0.64 & 0.64 \\ \hline
Black Box & 0.51 & 0.73 & 0.64 & 0.62 & 0.64 & 0.64 \\ \hline \hline
\multicolumn{7}{|c|}{\textbf{Average oracle TPR loss over all attacks}} \\
\hline \textbf{Scenario} & \textbf{FFNN} & \textbf{CANdito} & \textbf{Short LSTM} & \textbf{Long LSTM} & \textbf{Short GRU} & \textbf{Long GRU} \\ \hline
White Box & -20.63\% & -4.61\% & -23.97\% & -23.86\% & -21.31\% & -23.15\% \\ \hline
Grey Box & -6.02\% & 1.63\% & -7.99\% & -7.70\% & -7.15\% & -6.72\% \\ \hline
Black Box & -4.64\% & -0.19\% & -8.31\% & -8.63\% & -6.94\% & -6.58\% \\ \hline
\end{tabular}
}
\vspace{-10px}
\end{table}

\subsection{Experiment 2: Attack Perturbation}

\begin{figure}
    \centering
    \begin{subfigure}[b]{0.49\textwidth}
        \centering
        \includegraphics[width=\textwidth]{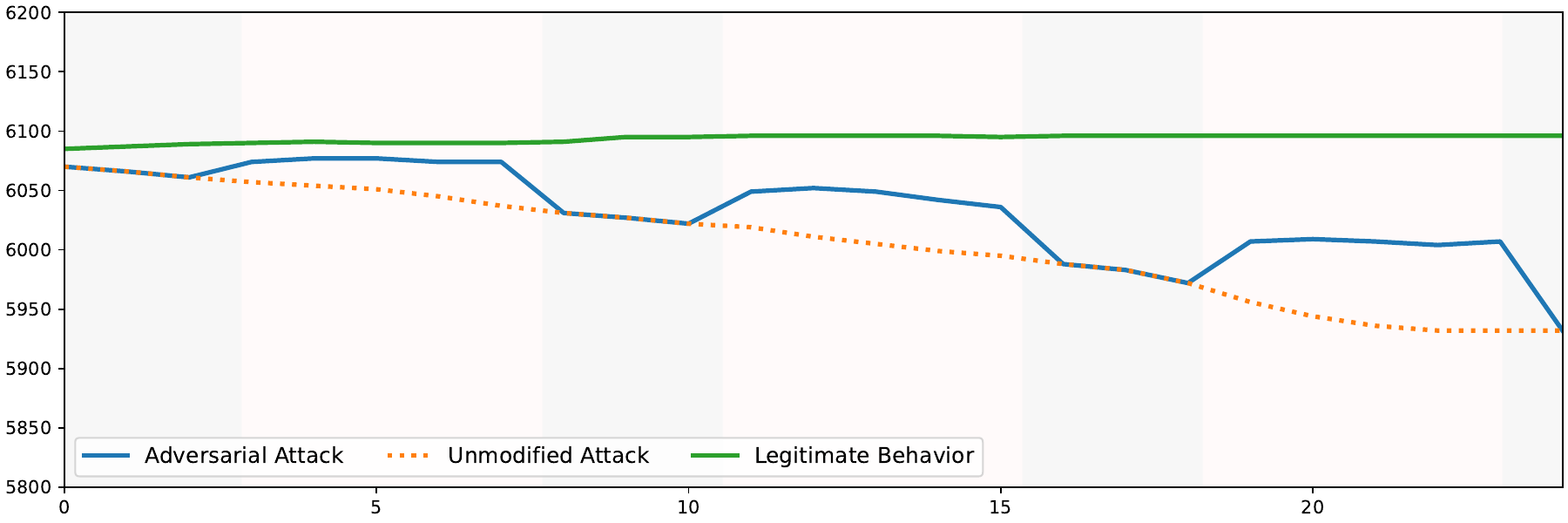}
        \caption{White-box, Short LSTM oracle, Masquerade replay attack, DeepFool evasion algorithm, ID 0DE.}
        \label{fig:subfig1}
    \end{subfigure}
    \hfill
    \begin{subfigure}[b]{0.49\textwidth}
        \centering
        \includegraphics[width=\textwidth]{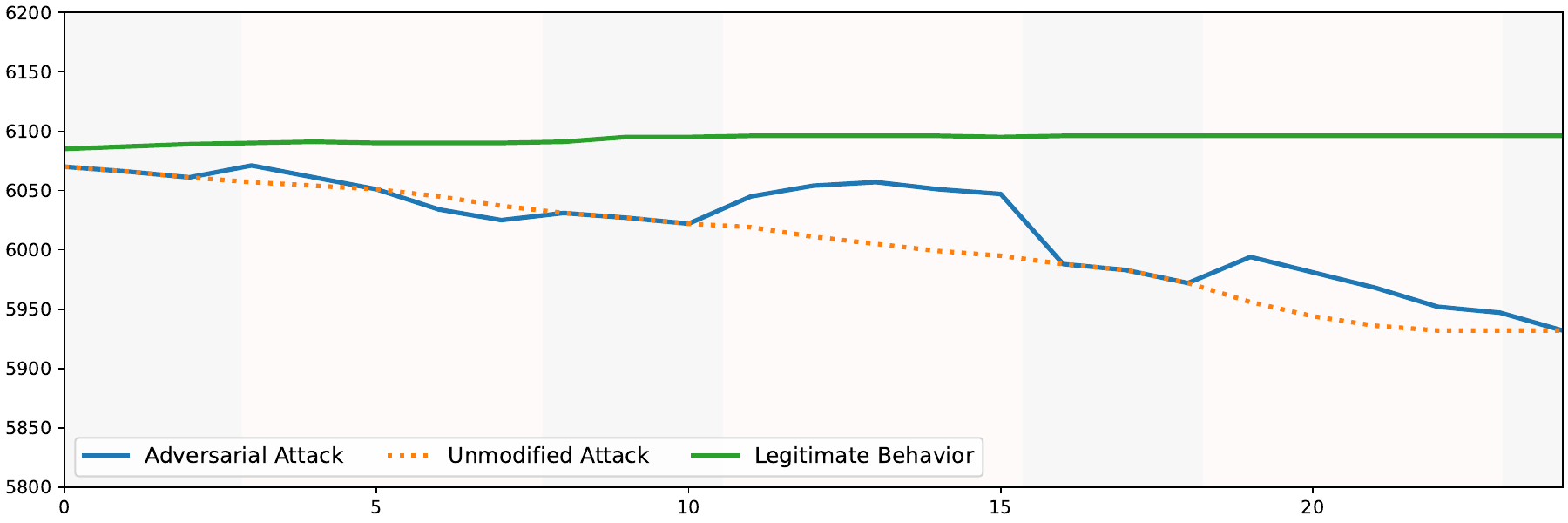}
        \caption{Black-box, Short LSTM oracle, Masquerade replay attack, DeepFool evasion algorithm, ID 0DE.}
        \label{fig:subfig2}
    \end{subfigure}
    
    \begin{subfigure}[b]{0.49\textwidth}
        \centering
        \includegraphics[width=\textwidth]{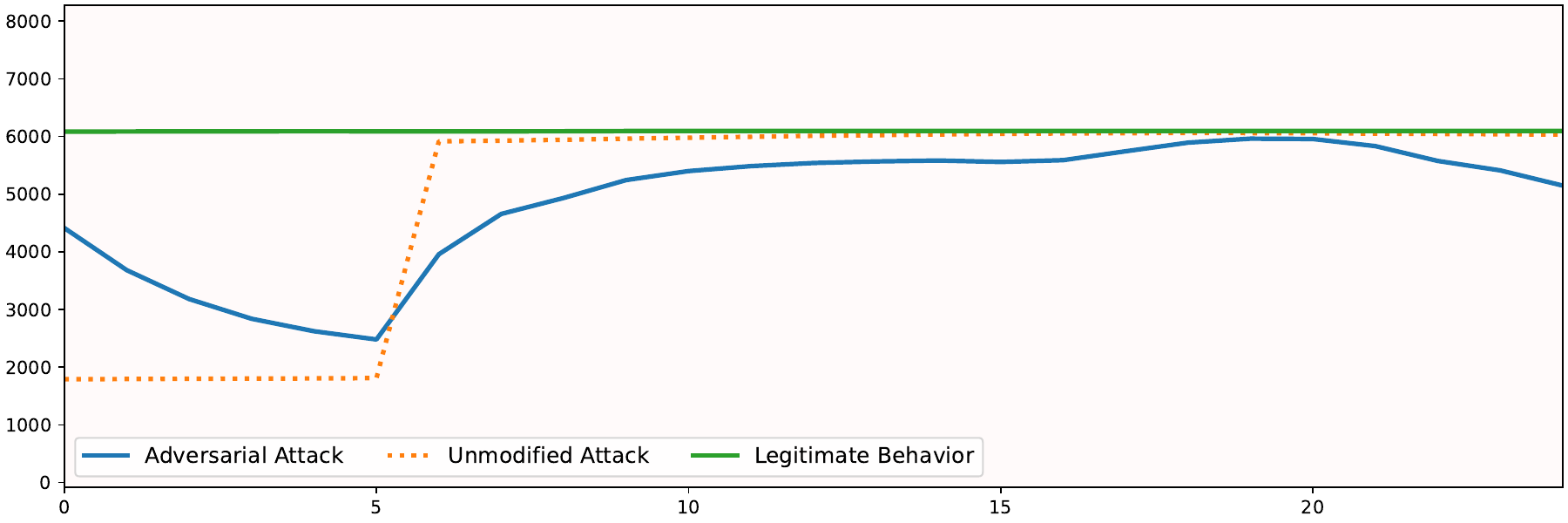}
        \caption{White-box, Short GRU oracle, Continuous change attack, DeepFool evasion algorithm, ID 0DE.}
        \label{fig:subfig3}
    \end{subfigure}
    \hfill
    \begin{subfigure}[b]{0.49\textwidth}
        \centering
        \includegraphics[width=\textwidth]{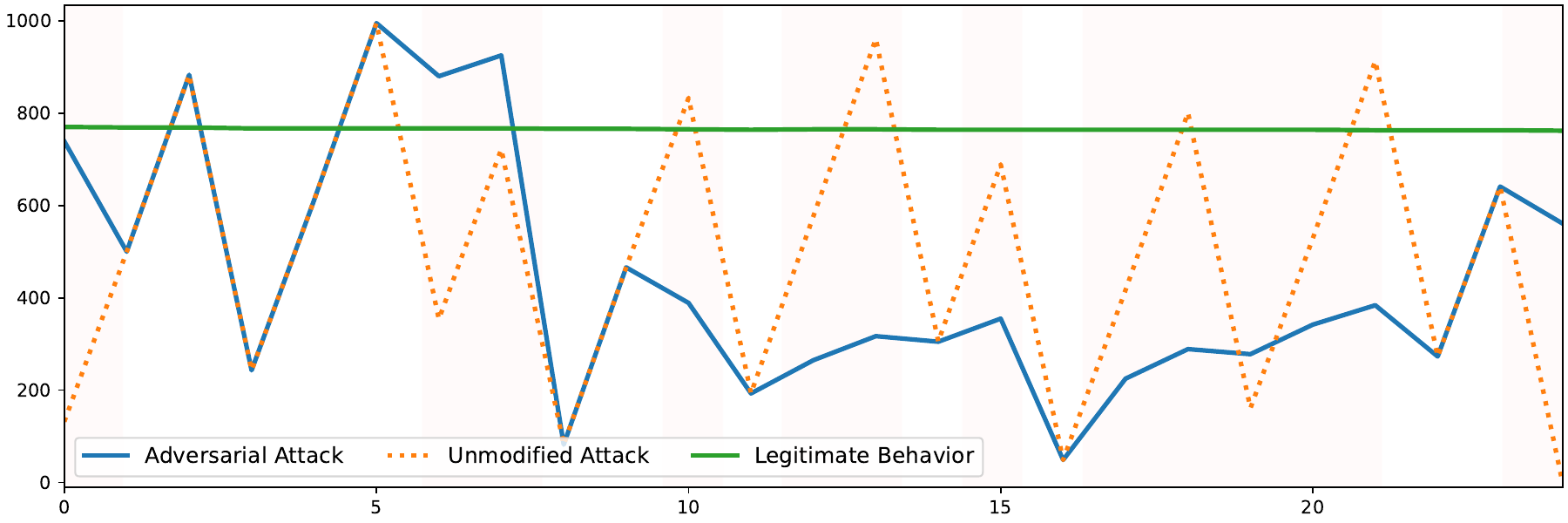}
        \caption{White-box, Short GRU oracle, Continuous change attack, DeepFool evasion algorithm, ID 0FF.}        
        \label{fig:subfig4}
    \end{subfigure}
    
    \begin{subfigure}[b]{0.49\textwidth}
\centering
        \includegraphics[width=\textwidth]{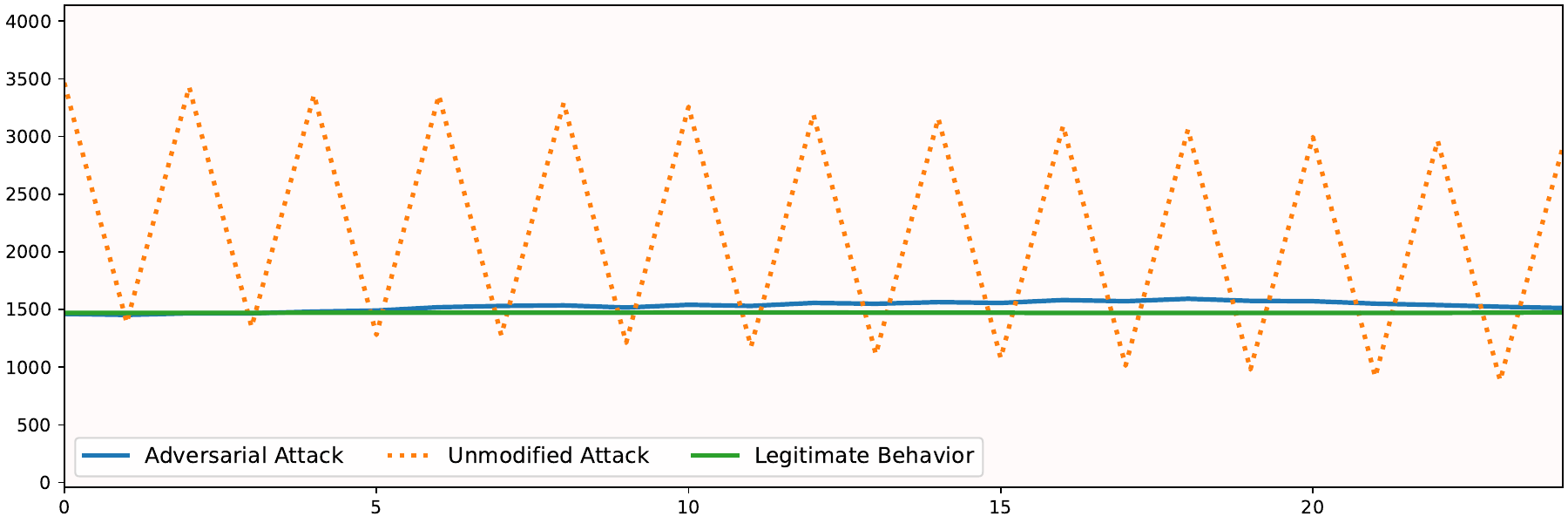}
        \caption{Black-box, Short GRU oracle, Continuous change attack, L2 BIM evasion algorithm, ID 11C.}
        \label{fig:subfig5}
    \end{subfigure}
    \hfill
    \begin{subfigure}[b]{0.49\textwidth}
        \centering
        \includegraphics[width=\textwidth]{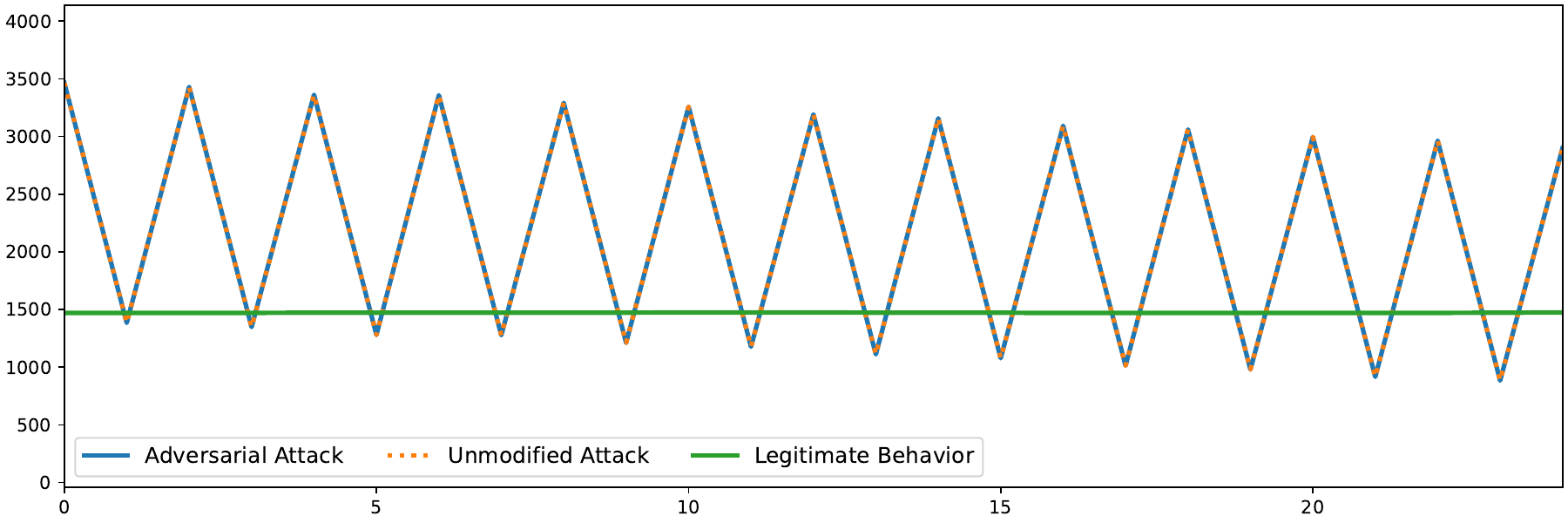}
        \caption{Black-box, Short GRU oracle, Continuous change attack, DeepFool evasion algorithm, ID 11C.}
        \label{fig:subfig6}
    \end{subfigure}
    
    \caption{Example plots of attack events in various scenarios. In each plot the green line shows the untampered legitimate behavior of the signal, the dotted yellow line shows the basic version of the attack, and the blue line shows the evasive version of the attack.}
    \label{fig:plots}
\end{figure}

Figure~\ref{fig:plots} shows some exemplary adversarial behaviors to provide a qualitative assessment of the effectivness of the adversarially perturbed attacks on the vehicle by comparing the shape of the original with the adversarial attacks signals In the plot. In the figures, the dotted lines represent the intended content of an injected sequence while the blue lines are the results of the adversarial perturbation and the green lines provide a baseline reference depicting the normal signal in the attack-free state; a red background highlights packets that have successfully evaded the \ac{IDS} and a grey background indicates the packets that were already undetected. We present three recurring scenarios worth discussing, cases in which the perturbation is effective, cases in which the perturbation nullifies the attacker goals, and cases in which the attack fails.

\mypar{Plots~\ref{fig:subfig1},~\ref{fig:subfig2}, and~\ref{fig:subfig3}} illustrate attack sequences that successfully achieve the initial attack goal while fully evading detection. It's worth noting, though unsurprising, that effective evasion strategies tend to involve staying close to the baseline reference (Plots~\ref{fig:subfig1} and~\ref{fig:subfig2}) and avoiding overly aggressive value changes (Plot~\ref{fig:subfig3}). Plots~\ref{fig:subfig1} and~\ref{fig:subfig2} depict nearly identical setups, with the key difference being the white-box vs. black-box scenario. Interestingly, the behavior of effective evasive sequences in both scenarios is often quite similar, as seen in this case. The main difference in evasion rate and perturbation, as previously shown in Table~\ref{tab:evasivealgorithmsperformances}, is typically due to the algorithm's failure to find an evasive perturbation, rather than any notable variation in the behavior of the generated evasive sequences themselves.

\mypar{Plot~\ref{fig:subfig5}} captures an output sequence of the continuous change attack against the short \ac{GRU} model for \ac{CAN} ID ``11C''. There are multiple instances, throughout the experiments, of attacks that evade in a similar pattern, where the evasive points are very close to the reference normal traffic that the \ac{ECU} would have transmitted if it was not silenced. This is particularly true for the two continuous experiments where the attack heavily manipulates only one specific signal and replays the others: the adversarial gradient ascent correctly captures the intra-packet dependency and pushes the rogue signal to values that are consistent with the context. This behavior is of course undesirable for a malicious actor since, despite the success of the evasion attempt, the meaning of the target payload is completely lost and the result is almost identical to not carrying out any attack. Note that not all evasive algorithms act similarly when attempting to generate an evasive attack, as visible in Plots~\ref{fig:subfig5} and~\ref{fig:subfig6}. Although both plots represent the same attack instance, scenario, and oracle, while the L2 BIM algorithm finds an evasive (but ineffective) perturbation, DeepFool fails.

\mypar{Finally, plots~\ref{fig:subfig4} and~\ref{fig:subfig6}} depict perturbation scenarios that can be considered failures. In Plot~\ref{fig:subfig4}, the attack signal's unpredictable behavior prevents the algorithm from identifying a suitable perturbation for several packets in the sequence. In most of the evaded instances, the perturbed signal does not reach values similar to the original attack, casting doubt on the attack's overall effectiveness. If we focus on the correctly flagged packets that the algorithm failed to modify (shown with a white background), it becomes evident that it is easier to perturb the signal toward values resembling the last received packet (note that this behavior was not seen in CANdito, as the initial fully connected layer and target sequence reversal prevent the model from overly weighting the most recent packets during reconstruction, while also using a non-overlapping sliding window input). In Plot~\ref{fig:subfig6}, the algorithm completely fails to find a single evasive point throughout the entire sequence. It is worth noting that the algorithm attempts to find an evasive point for up to 50 steps, and depending on the attack and oracle, some algorithms may have slower or more aggressive modification curves, which could explain the significant differences compared to Plot~\ref{fig:subfig5}.

\subsection{Experiment 3: Attack Precomputation}
\label{exp:precomputation}

The strategy described to generate the adversarial attacks requires repeated querying of a target autoencoder or predictor. This does not fit well with the speed of the CAN bus signals, therefore we test whether an attacker could compute a sequence of adversarial packets in advance and successfully inject it at a later time while avoiding detection. In practice, this experiment takes all those sequences that are \emph{fully evasive}, i.e., exclusively composed of packets classified as normal traffic, and tries to find similar points in the flow of messages where they could equally evade the \ac{IDS}. We also consider as candidate injection points every point in the traffic preceded by at least ten packets identical to the preamble found at the original attack location. We exclude from this test the FFNN model, as it performs classification independently of the order or position of messages, and CANdito as, given its superior resilience, there were not enough completely evasive sequences to carry out the test.  Moreover, we provide this evaluation only for the ReCAN dataset, since the attacks on the CarHacking dataset are per-se injection attacks, and would anyway disrupt the normal flow of the network.
The results are widely dependent on the specific \ac{CAN} ID, with two clear clusters: 

    \mypar{Cluster 1} IDs ``1FB'' and ``104'' have very slowly varying physval signals, that for the duration of the traffic gravitate around some common values rather than assuming any possible bit configuration (albeit all configurations are valid and no bit always remains constant). This behavior causes a relatively high number of possible reinjection points, with a peak cardinality of nearly 1800 points found for a single sequence, and a high success rate. In general, sequences similar to the one in the view of Plot~\ref{fig:subfig1} transfer easily into spots in the traffic with a longer matching preamble, as we observe a $95\%$ success rate with an average number of 38 identical preceding packets, obtaining several hundreds of potentially \emph{fully evasive} sequences from 4 to 10 precomputed attacks, depending on the ID, attack, and oracle;
    
    \mypar{Cluster 2} The remaining 10 IDs do not bring the same degree of success as they provide way fewer injection points, with many preambles without a match in the whole traffic flow; once again, the few successful precomputed attacks require a preamble almost identical to the original, with over 37 matching messages on average.
In general, this precomputed attack is feasible and reliable only within the trivial case of a preamble that is close to the original \ac{IDS} input window, with more of the $90\%$ of the successful reinjections differing only for a couple of packets. This makes just a few specific devices among the considered \acp{ECU} vulnerable to the approach under testing, however, it is not possible to ascertain the impact of the resulting risk with the current information about the function and semantic associated with each affected \ac{CAN} ID.

\subsection{Discussion on Defenses and Mitigation Strategies}
\label{sec:defenses}

While the focus of this work is on evaluating the vulnerabilities of machine learning-based intrusion detection systems (IDSs) in automotive networks, it is important to discuss potential mitigation strategies. The two main directions we envision are adversarial training and input preprocessing:

Adversarial training~\cite{goodfellow2015explaining} involves incorporating adversarial examples during model training to improve robustness. This technique has shown promise in other domains such as computer vision, and could, in principle, be adapted to tabular and temporal domains like CAN traffic. However, applying adversarial training in the automotive setting presents unique challenges. In particular, introducing perturbed sequences during training may shift the model’s decision boundary in a way that increases false positives. Given the safety-critical nature of automotive systems, an increase in false alarms may render an IDS practically unusable.

Input preprocessing techniques (e.g., smoothing, quantization, or feature selection) have been proposed in other fields~\cite{guo2017countering} to reduce the impact of adversarial noise. Some of these approaches could be adapted to the CAN context, especially given its structured and discrete nature. However, their integration requires engineering to avoid excluding critical information from the detection process, which would allow the attacker to bypass the detection system by injecting attacks that do not affect the chosen features.

\section{Conclusions}
\label{sec:conclusion}

In this paper, we addressed the impact of adversarial attacks on state-of-the-art automotive \acp{IDS}. We conducted a thorough evaluation of known adversarial evasion attacks, adapted to the automotive domain, against payload-based \acp{IDS} using real CAN traffic over two public datasets. We designed and implemented customized variants of the popular \ac{BIM} and DeepFool perturbation algorithms, and tested six different detection architectures from the state of the art in white-, grey-, and black-box scenarios. Our results show that evasion is achievable, although not always consistently, especially in the white-box scenario, across both datasets. In the grey- and black-box scenarios, the performance degradation of the intrusion detection systems is notably reduced. DeepFool proves to be the most effective evasion algorithm, generating numerous evasion points on both datasets and achieving up to an 85\% drop in TPR. 
Exploring the characteristics of evasion samples, we observe that in multiple instances, either part of the attack sequence fails to evade detection or the perturbation effects render the payload similar to a non-tampered one. We evaluate the feasibility of precomputing the evasive sequence and injecting it, demonstrating that this is often possible.

The most significant limitation of our study stems from the lack of availability of a real test vehicle, where the injected packets may influence the vehicle's and bus's behavior—a characteristic that cannot be evaluated with previously collected traffic logs. Future work will aim to evaluate how well perturbed attack sequences preserve the original attack intent, providing a more comprehensive understanding of evasion algorithms—not only in terms of bypassing detection but also in achieving the attacker’s intended impact. Additionally, further research should investigate the applicability and effectiveness of alternative adversarial approaches, such as score-based methods, in tabular and temporal domains like those found in automotive systems.

\begin{acks}
This work was partially supported by projects SERICS (PE00000014) under the NRRP MUR program funded by the EU - NGEU and MICS (Made in Italy – Circular and Sustainable) from Next-Generation EU. CUP MICS D43C22003120001.
\end{acks}

\bibliographystyle{ACM-Reference-Format}
\bibliography{bib}










\end{document}